\documentclass[onefignum,onetabnum]{siamart190516}

\usepackage{lipsum}
\usepackage{amsfonts}
\usepackage{algorithmic}
\usepackage{amssymb}
\usepackage{amsmath}
\usepackage{graphicx}
\usepackage{subfig}
\usepackage{epstopdf}
\usepackage{bm}
\DeclareMathOperator{\sgn}{sgn}

\ifpdf
  \DeclareGraphicsExtensions{.eps,.pdf,.png,.jpg}
\else
  \DeclareGraphicsExtensions{.eps}
\fi

\newsiamremark{remark}{Remark}
\newsiamremark{hypothesis}{Hypothesis}
\crefname{hypothesis}{Hypothesis}{Hypotheses}
\newsiamthm{claim}{Claim}

\headers{Disconnection Model for Triple Junction}{C. Wei, L. Zhang, J. Han, D. J. Srolovitz and Y. Xiang}

\title{Grain boundary triple junction dynamics: \\a continuum disconnection model
}

\author{Chaozhen Wei\footnotemark[2] \footnotemark[3]
\and Luchan Zhang\footnotemark[2]
\and Jian Han\footnotemark[4] \footnotemark[6]
\and David J. Srolovitz\footnotemark[4] \footnotemark[5] \footnotemark[6]
\and Yang Xiang\footnotemark[2]
}

\usepackage{amsopn}

\begin{document}

\maketitle

\renewcommand{\thefootnote}{\fnsymbol{footnote}}

\footnotetext[2]{Department of Mathematics, The Hong Kong University of Science and Technology, Clear Water Bay, Kowloon, Hong Kong SAR, China
  (\email{maxiang@ust.hk}).}
\footnotetext[3]{Institute for Advanced Study, The Hong Kong University of Science and Technology, Clear Water Bay, Kowloon, Hong Kong SAR, China (\email{iasczwei@ust.hk}).}
\footnotetext[4]{Department of Materials Science and Engineering, University of Pennsylvania, Philadelphia, PA 19104-6272 USA.}
\footnotetext[5]{Department of Mechanical Engineering and Applied Mechanics, University of Pennsylvania, Philadelphia, PA 19104-6272 USA.}
\footnotetext[4]{Department of Materials Science and Engineering, City University of Hong Kong, Hong Kong SAR, China (\email{srol@cityu.edu.hk}).}

\renewcommand{\thefootnote}{\arabic{footnote}}

\begin{abstract}
The microstructure of polycrystalline materials consists of networks of grain boundaries (GBs) and triple junctions (TJs), along which three GBs meet.
The evolution of such microstructures may be driven by surface tension (capillarity), applied stresses, or other means that lead to a jump in chemical potential across the GBs.
Here, we develop a model for the concurrent evolution of the GB/TJ network based upon the microscopic mechanism of motion; the motion of line defects (disconnections) in the GB that have  both dislocation and step character.
The evolution involves thermally-activated disconnection formation/annihilation and migration of multiple disconnections modes/types.
We propose this crystallography-respecting continuum model for the disconnection mechanism of GB/TJ dynamics derived with a variational approach based on the principle of maximum energy dissipation.
The resultant TJ dynamics is reduced to an optimization problem with  constraints that account for local microstructure geometry, conservation of Burgers vectors, and thermal-kinetic limitations on disconnection fluxes.
We present analysis of and numerical simulations based upon our model to demonstrate the dependence of the GB and TJ mobilities and the TJ drag effect on the disconnection properties, and compare the predictions with molecular dynamics and experimental observations.
\end{abstract}



\section{Introduction}
\label{sec:intro}
A polycrystalline material consists of domains (grains) that share the same crystal structure but of different crystal orientations. Grain Boundaries (GBs) are domain walls across which the orientation changes (typically atomically sharp).
Three grains and three GBs meet at a Triple Junction (TJ) and four grains, six GBs and four TJs meet at a Quadruple Point (QP) -- as required by Euler's formula.
GBs are co-dimension 1 objects, TJs are co-dimension 2 objects, and QPs are co-dimension 3 objects; i.e., in a 3-dimensional space, GBs are planes, TJs are lines, and QPs are points.
Polycrystals may be viewed, alternatively, as networks of GBs or TJs.
The scale and/or geometry of polycrystalline microstructures strongly affect many mechanical, thermal and electronic properties of polycrystalline materials \cite{sutton1995interfaces}.

In normal grain growth, microstructures evolve such that the energy of these networks decreases (mean grain size increases).
This evolution involves the concurrent motion of GBs, TJs and QPs. In the limit that all GBs have the same energy and mobility, all motion is overdamped, and junctions have infinite mobility, the microstructure evolution reduces to GB mean-curvature flow. In this limit, the evolution of the size of any grain in the network is known exactly in any number of dimensions; in 2 dimensions the prediction depends only on grain topology \cite{von1952metal,Mullins1956two}, while in higher dimensions it is also a function of grain shape \cite{macpherson2007neumann}.

Conventional grain growth theory \cite{Mullins1956two,hillert1965on} describes GB migration as mean-curvature flow in the direction normal to the GB. In this theory, TJs are assumed to not disturb grain boundary migration but simply follow the motion of the constituent grain boundaries; TJs move to maintain the balance of the surface tensions of their three constituent GBs.
As a result, triple junctions migrate with a set of fixed dihedral angles that satisfy a local equilibrium condition, i.e., the Herring equation \cite{herring1951surface,King1999}.
This TJ dihedral angle equilibrium condition has long  served as a fundamental boundary condition for surface tension/capillarity-driven GB migration in grain growth theories \cite{von1952metal,macpherson2007neumann} and computer simulations \cite{kinderlehrer2001evolution,Kinderlehrer2006variation,elsey2009diffusion,lazar2010accurate,esedoglu2016grain} as well as in interpretation of experimental observations \cite{adams1999extract,morawiec20003method}.

Conventional grain growth theory rests on an implicit assumption that TJ mobility is infinite; i.e., TJs can always migrate sufficiently fast to remain in equilibrium with the GBs and the equilibrium TJ dihedral angles are always maintained.
There is, however, substantial experimental \cite{galina1987influence,CZUBAYKO1998influence,gottstein1999TJ} and atomic simulation \cite{Upmanyu1999TJMD,Upmanyu2002TJMD} evidence that demonstrates that this assumption is often not true.
Finite TJ mobility manifests itself in the form of a drag effect that leads to the deviation of dynamical dihedral angles from their thermodynamic equilibrium values (as set by the Herring relation) \cite{galina1987influence,CZUBAYKO1998influence,gottstein1999TJ,Mattissen2005TJdrag,Upmanyu1999TJMD}.
Assuming that  TJ dynamics is overdamped (i.e., with a finite TJ mobility), the TJ  velocity is proportional to the driving force resulting from the net surface tension from its constituent GBs.
Hence, TJ motion is a kinetically-limited dynamic process; the TJ moves toward the location at which the dihedral angles are in equilibrium with a finite velocity driven by the imbalanced surface tensions at the TJ.
The TJ drag effect has been widely investigated in steady-state TJ migration (with carefully tailored GB profiles ) \cite{CZUBAYKO1998influence,gottstein2002TJ,Upmanyu2002TJMD}.
A recent analysis \cite{ZHAO2017345} also examined  the effect of non-steady-state TJ motion following topological changes in a polycrystalline microstructure.
Incorporation of the TJ drag effect into the otherwise classical von Neumann-Mullins relation \cite{von1952metal} modifies the classical grain growth prediction of the evolution of mean grain size with time \cite{gottstein2002TJ}.
The transition between the GB migration-dominated grain growth kinetics and TJ mobility-limited kinetics occurs at small grain sizes (where driving forces are large) \cite{gottstein1999TJ,Upmanyu2002TJMD,johnson2014phasefield} and/or at low temperatures (where the TJ mobility is low) \cite{CZUBAYKO1998influence,Upmanyu2002TJMD,Mattissen2005TJdrag}.
While these analyses are able to account for the effects of finite TJ mobility on microstructure evolution, the underlying mechanisms that control TJ mobility remain unknown and unaccounted for; hence, the effects of finite TJ mobility on GB migration remain, at best, heuristic.

It is now widely accepted that GB migration in many polycrystalline materials is mediated/controlled by line defects that are constrained to move within the GB.
These defects, known as disconnections, are characterized by both a Burgers vector $\bm{b}$ (dislocation character) and a step height $H$ (step character).
The glide motion of disconnections along the GB leads to GB migration in the direction of the GB normal (associated with the step character) and shear or tangential translation of one grain with respect to the other  (associated with the Burgers vector) \cite{Rajabzadeh2013evidence,rajabzadeh2013elementary}.
The disconnection character is limited by the underlying GB crystallography \cite{bollmann1970crystal,hirth1996disconnection}.
More specifically, GB bicrystallography allows for an infinite, discrete set of possible disconnection types or modes $\{\bm{b},H\}$ \cite{han2018}; the relative importance of these modes depends on disconnection mobilities and equilibrium disconnection densities,  and may be analyzed within a statistical mechanics framework \cite{thomas2017reconciling,chen2019shear}.
Many GB dynamical phenomena require consideration of multiple disconnection modes \cite{han2018}.
A continuum framework for grain boundary migration that accounts for the underlying disconnection dynamics was recently proposed \cite{zhang2017prl,wei2019gb_mm}. In addition, a continuum model for low-angle grain boundaries incorporating dislocation structure was also proposed \cite{zhang2018lowangle}.

Despite the recent progress in describing GB dynamics based on the underlying disconnection motion, a disconnection-based understanding of triple junction motion is only now developing \cite{han2018,thomas2019TJ,zhu2019situ}.
Since a GB is inevitably delimited by triple junctions in a polycrystal, triple junctions serve as obstacles to the motion of disconnections along GBs; a disconnection cannot propagate from one GB to another by simply traversing the TJ since the disconnection modes ($\bm{b},H$) in different GBs are, in general, not the same.
The disconnection flux into TJs are sources of stress and Burgers vector accumulation at TJs that will result in the stagnation of GB migration \cite{thomas2019TJ} (the stress field associated with the accumulated Burgers vector at the TJs limits disconnection nucleation/migration on the GBs).
Conversely, triple junctions may also serve as sources for defects such as disconnections \cite{zhu2019situ}, lattice dislocations \cite{Hashimoto1987,Chen2003science} and twins \cite{thomas2016twin,lin2015}.
Such disconnection formation at the TJs and their subsequent motion along the GBs  leads to concurrent TJ and GB migration \cite{zhu2019situ}.

A limited continuum description of  TJ motion based upon disconnection dynamics  was first proposed in \cite{zhang2017prl}.
This model connects TJ motion with the disconnection flux into and the reactions at the TJ through  the conservation of both disconnection Burgers vector and step height.
This original continuum, disconnection model for TJs was applied (and restricted) to a highly idealized GB/TJ geometry.
However, several important issues remain:
(1) TJ motion described in the original model is not  consistent with the migration of general GBs.
(2) The flux of disconnection Burgers vectors into a TJ from the constituent GBs generally results in Burgers vector accumulation at the TJ which generates long range stresses that affect disconnection mode selection and GB/TJ evolution.
(3) The ``apparent" or ``effective'' GB mobility is a statistical average of the properties of the multiple disconnection modes that are operative and their responses to different types of driving forces \cite{chen2019shear,wei2019gb_mm}.
(4) The relation between the disconnection mechanism-based description of TJ motion and the conventional capillarity-driven model is unclear.
Recent molecular dynamics simulations of TJ migration provided some insight into TJ motion that was used to develop a less restrictive continuum model for TJ migration \cite{thomas2019TJ}, but was still based on the assumption of zero Burgers vector accumulation.

Here, we propose a continuum model (in two dimensions) for coupled GB/TJ migration based on the disconnection mechanism that provides a geometrically consistent description of GB/TJ evolution and explicitly accounts for Burgers vector accumulation and relaxation at TJs.
We provide a detailed derivation of this continuum model by a variational approach based on the principle of maximum dissipation \cite{Svoboda2005pmd,Hackl2008pmd}, which is equivalent to the Onsager's variational principle by minimizing the so-called Rayleighian function. This variational approach has been used for formulating evolution equations in fluid dynamics (e.g., \cite{qian2006}), soft matter physics (e.g., \cite{Doi2011}) and solid mechanics (e.g., \cite{cocks1996,fisher2012,Liu2019gb,zhang2018lowangle} for capillarity-driven GB dynamics).
The resultant model for TJ migration is formulated as an optimization problem with the necessary constraints that account for GB and TJ crystallography.

The remainder of this paper is organized as follows. In Section~\ref{sec:capillary_model}, we briefly review the conventional capillarity-driven model for the GB/TJ migration.
In Section~\ref{sec:GB_model}, we introduce the disconnection mechanism for GB migration and present a continuum model of GB migration based on multiple disconnection modes \cite{wei2019gb_mm}.
In Section~\ref{sec:TJ_model}, we derive a disconnection mechanism-specific continuum model for  coupled GB/TJ migration based upon the principle of maximum dissipation.
In Section~\ref{sec:simulation}, we perform analytic analysis and numerical simulations of coupled GB/TJ migration.

\section{Capillarity-driven microstructure evolution}
\label{sec:capillary_model}
The conventional equation of motion for capillarity-driven GB migration assumes that GB migration is overdamped, such that the GB velocity is proportional to the driving force (i.e., negative first variation of the free energy with respect to the displacement of a GB along its normal).
The total free energy is the integral of the GB energy $\gamma(\chi)$ (which may be a function of GB inclination $\chi$) over all of the GBs in the network.
This yields
\begin{equation}\label{eq:GB_capi}
\bm{v}_{GB} = M_{GB} \Big(\gamma+\frac{d^2\gamma}{d\chi^2}\Big)\kappa \bm{\eta},
\end{equation}
where $M_{GB}$ is the GB mobility, $\kappa$ is the GB mean curvature and $\bm{\eta}$ is the unit vector of the GB local normal.
The term $(\gamma+d^2\gamma/d\chi^2)$ is the surface stiffness first proposed by Herring \cite{Herring1999}.
If $\gamma$ is independent of GB inclination, then Eq.~(\ref{eq:GB_capi}) reduces to mean curvature flow.

\begin{figure}[ht]
\centering
\subfloat[\label{fig:TJ_circle}]{\includegraphics[width=0.33\textwidth]{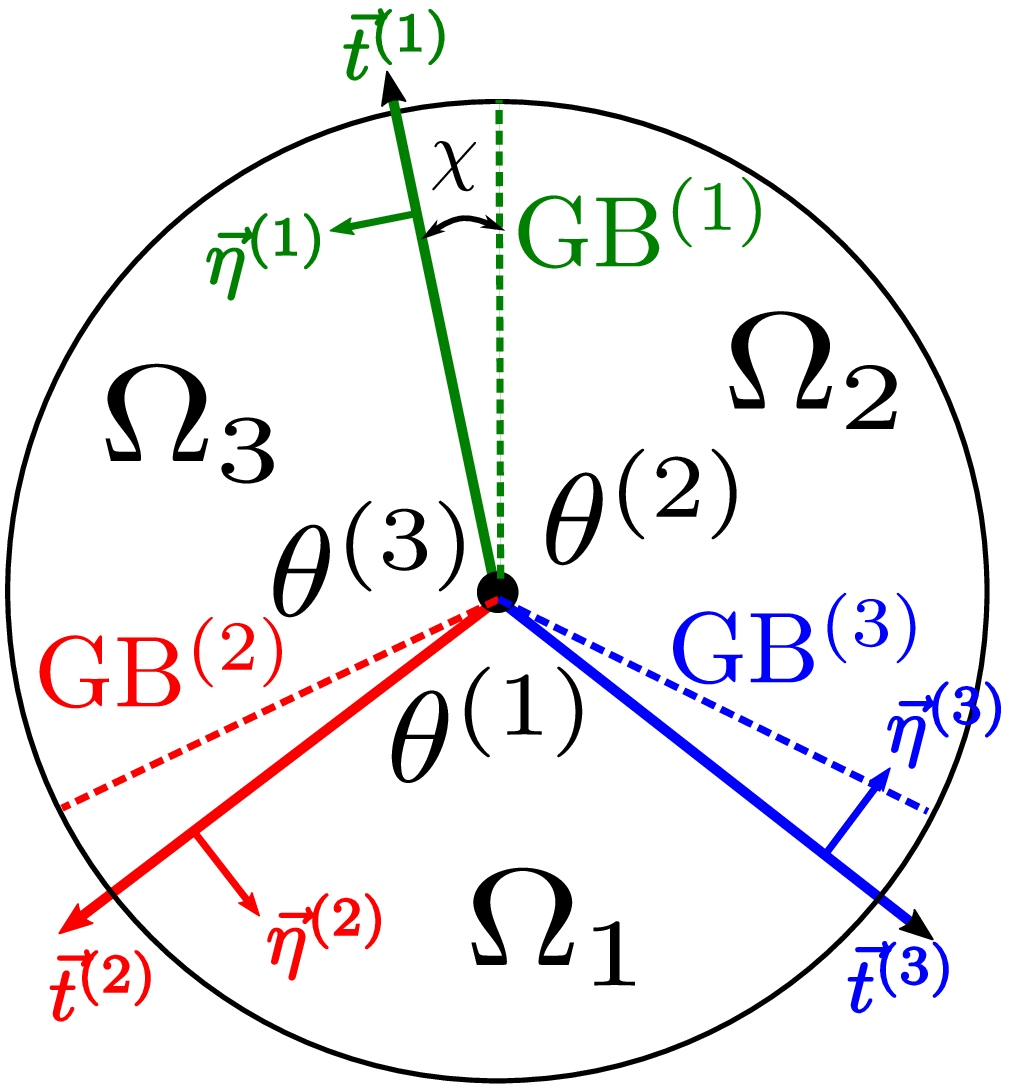}}\hspace{5pt}
\subfloat[\label{fig:TJ_velocity}]{\includegraphics[width=0.43\textwidth]{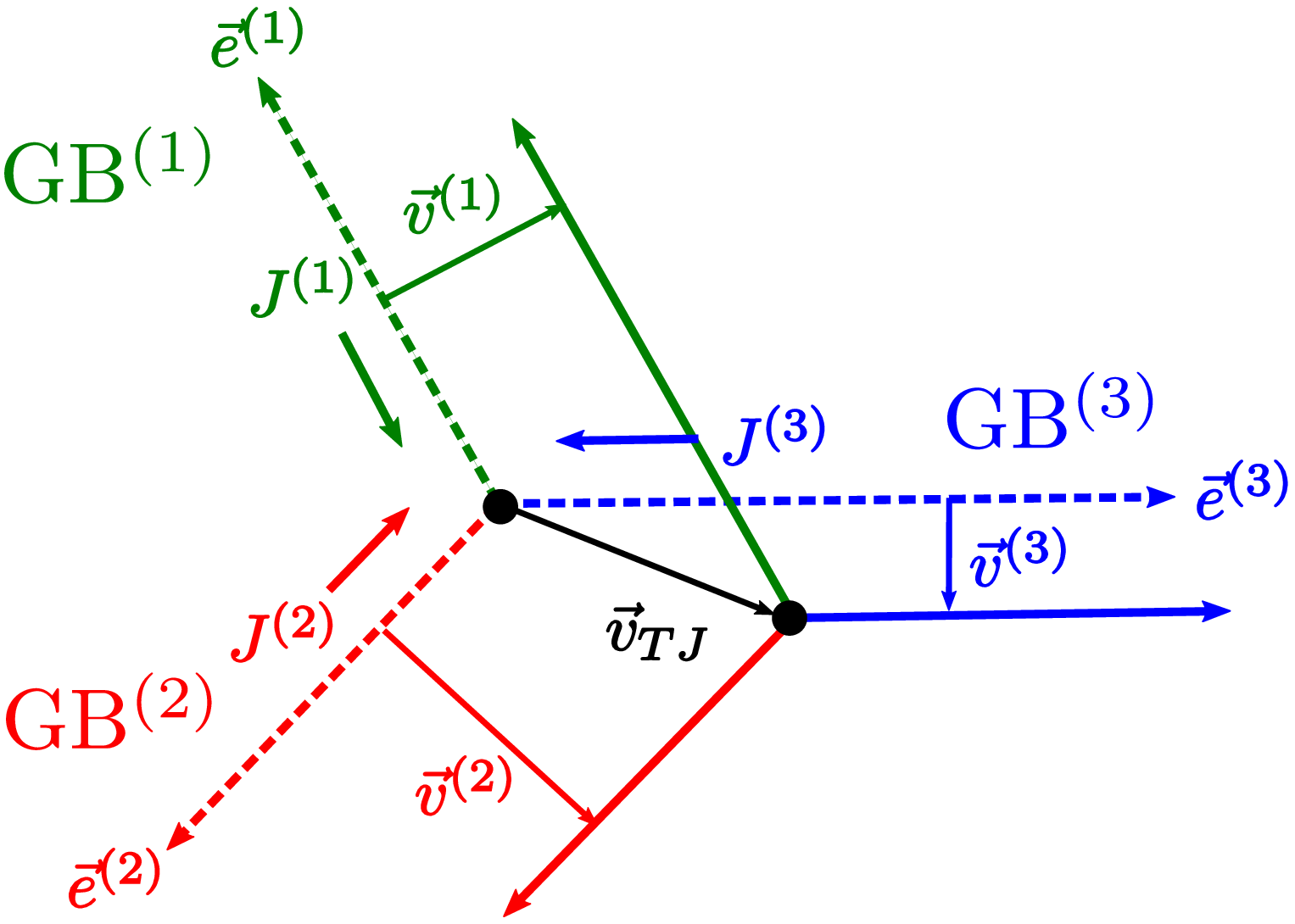}}\hspace{5pt}
\subfloat[\label{fig:TJ_angles}]{\includegraphics[width=0.17\textwidth]{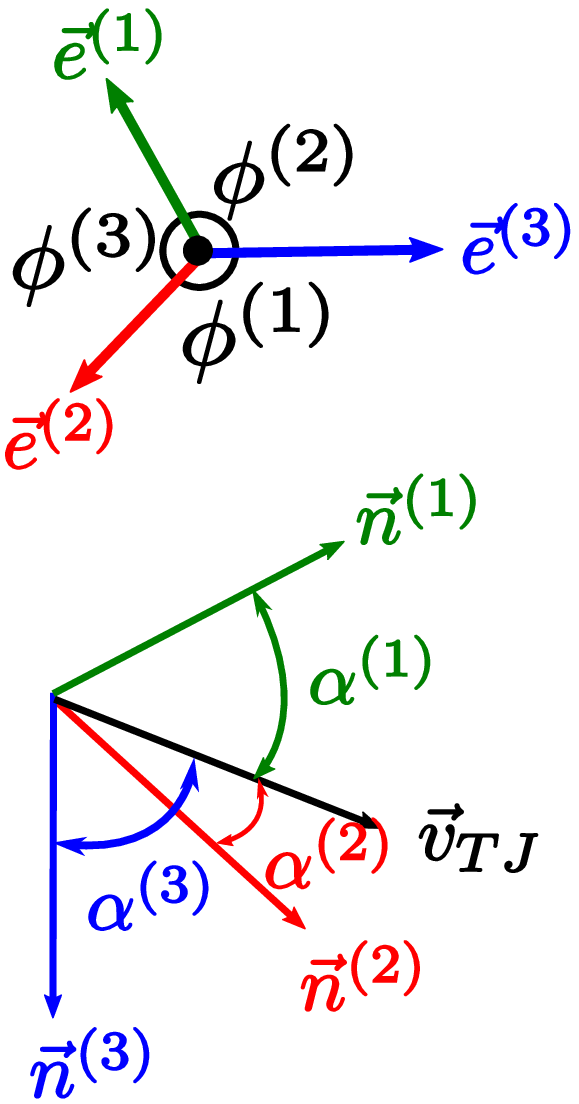}}
\caption{(a) A schematic  of a section of a 2D polycrystal of three grains (i.e., $\Omega_I$) consisting of a triple junction and its three constituent grain boundaries.
Solid lines (radial vectors) represent the current GB inclinations; dashed lines represent the reference inclinations (i.e., $\bm{e}^{(j)}$) of the GBs.
$\bm{t}^{(j)}$ is the unit tangent to GB$^{(j)}$ (current inclination), $\bm{\eta}^{(j)}$ is its unit normal, $\theta^{(j)}$ is the TJ dihedral angle opposite to GB$^{(j)}$, and $\chi$ is the GB inclination angle between the current and reference inclinations measured in the counterclockwise sense.
(b) A schematic of the disconnection mechanism for TJ motion via disconnection fluxes $J^{(j)}$ on the constituent GBs, showing the relationship between the TJ velocity $\bm{v}_{TJ}$ and the GB velocities $\bm{v}^{(j)}_{GB}$.
The initial/final GB positions are shown as dotted
/solid lines.
(c) Schematic illustration of the dihedral angles $\phi^{(j)}$ between the reference GB planes $\bm{e}^{(j)}$ (parallel to the disconnection Burgers vector), and the angles $\alpha^{(j)}$ between the TJ velocity $\bm{v}_{TJ}$ and the GB velocities $\bm{v}^{(j)}$ in the directions normal to the reference GB plane, $\bm{n}^{(j)}$.
}\label{fig:TJ_fig}
\end{figure}

For a triple junction at equilibrium, the surface tensions from the three constituent GBs balance at the TJ; this gives the Herring equation
\begin{equation}\label{eq:herring}
\sum_{j=1}^3 \Big(\gamma^{(j)} \bm{t}^{(j)}+\frac{d\gamma^{(j)}}{d\chi}\bm{\eta}^{(j)}\Big) =\bm{0},
\end{equation}
where $\gamma^{(j)}$ is the GB energy of the $j^{\text{th}}$ constituent GB, $\bm{t}^{(j)}$ and $\bm{\eta}^{(j)}$ are the tangential and normal unit vectors of the $j^{\text{th}}$ GB at the TJ, respectively (see Fig.~\ref{fig:TJ_circle}). The derivative term is the torque that accounts for the effect of the inclination-dependence of the GB energy.
For the special case where the GB energy is inclination-independent, i.e., $\gamma^{(j)}(\chi)=\gamma^{(j)}$, the equilibrium condition for the dihedral angles reduces to Young's equation
\begin{equation}
\frac{\gamma^{(1)}}{\sin\theta^{(1)}}=\frac{\gamma^{(2)}}{\sin\theta^{(2)}}=\frac{\gamma^{(3)}}{\sin\theta^{(3)}},
\end{equation}
where $\theta^{(j)}$ is the dihedral angle opposite the $j^{\text{th}}$ GB (see Fig.~\ref{fig:TJ_circle}).

If the TJ moves sufficiently fast (relative to the GB migration), the TJ dihedral angles can quickly adjust via TJ migration when the GBs move such that they are always at their  equilibrium values during the GB/TJ network evolution.
The resultant grain growth is governed by the capillarity-driven grain boundary migration with the equilibrium boundary condition of dihedral angles at TJs, leading to the classical von Neumann-Mullins relation for grain size evolution in two-dimensions \cite{von1952metal,Mullins1956two} and the MacPherson-Srolovitz formula extending it to all dimensions $\geq2$ \cite{macpherson2007neumann}. However, if the TJ mobility $M_{TJ}$ is finite, the TJs cannot, in general, maintain the equilibrium dihedral angles during grain boundary migration.
Rather, the TJ exerts a drag on  GB migration and, hence, slows  grain growth.
Assuming that the TJ motion is overdamped, the TJ velocity is
\begin{equation}\label{eq:TJ_capi}
\bm{v}_{TJ} = M_{TJ}\sum_j \Big(\gamma^{(j)}\bm{t}^{(j)}+\frac{d\gamma^{(j)}}{d\chi}\bm{\eta}^{(j)}\Big).
\end{equation}

The TJ drag effect on  grain growth may be incorporated into a modified von Neumann-Mullins relation \cite{gottstein2002TJ}.
In this  model, the TJ drag effect is controlled by the dimensionless parameter
\begin{equation}\label{eq:drag_delta}
\delta =wM_{TJ}/M_{GB},
\end{equation}
where $w$ is a characteristic length (e.g., the average grain size); note that the ratio $M_{GB}/M_{TJ}$ has the dimensions of length.
As $\delta\rightarrow \infty$, the TJ is able to move quickly such that the equilibrium dihedral angles are always maintained (i.e., no TJ drag) and  grain growth is controlled by grain boundary migration.
On the other hand, as $\delta\rightarrow 0$, all GBs become flat and  grain growth is  controlled by  TJ dynamics.
For intermediate values of  $\delta$,  grain growth is a compromise between GB migration- and TJ drag-control.
The parameter $\delta$ is  small for small grain size \cite{gottstein1999TJ,Upmanyu2002TJMD,johnson2014phasefield}, low temperature \cite{CZUBAYKO1998influence,Upmanyu2002TJMD,Mattissen2005TJdrag}, and GBs with (or near) low-$\Sigma$ misorientations \cite{Upmanyu1999TJMD,Upmanyu2002TJMD}, as demonstrated in experiments and atomistic simulations such that TJ drag  is significant.

While both  GB and TJ mobilities are intrinsic properties of these defects and are sensitive to local bonding and local (atomic-scale) structure, they are constrained by  the underlying crystallography.
The relation between GB and TJ mobility remains unknown -- both theoretically and experimentally (except in a very small number of special cases).
The theory of capillarity-driven grain growth is incomplete in the sense that it provides no information on the underlying parameters ($\gamma$, $M_{GB}$, $M_{TJ}$) and their dependence on the macroscopic degrees of freedom (temperature, relative grain orientations, GB inclination) \cite{thomas2017reconciling,thomas2019TJ}.
In this paper, we present a crystallography-respecting continuum model for the evolution of the GB/TJ network based on the underlying disconnection mechanism.

\section{Disconnection model for grain boundary motion}
\label{sec:GB_model}
Continuum models for  disconnection-mediated GB migration based  on a single  \cite{zhang2017prl} and on multiple disconnection modes \cite{wei2019gb_mm} were previously proposed.
Consider a GB inclined relative to a high-symmetry GB inclination (usually that  for which the GB is a mirror plane).
In the two-dimensional case, we denote this reference inclination as $x$ and describe the GB shape in terms of its height measured relative to the $x$ axis, i.e., $y=h(x,t)$.
Suppose that there are multiple disconnection modes along the GB  each described by $(\bm{b}_i, H_i)$ (consistent with the underlying bicrystallography), where $i=1,2, \ldots$ is a disconnection mode index.
Assuming that the GB is close to the reference inclination (i.e., $|h_x| \ll 1$), the Burgers vector is nearly parallel to the GB plane such that $\bm{b}_i=(b_i,0)$ is along $x$-axis and $H_i$ is along $y$-axis.
In a discrete picture, the GB is composed of an array of disconnections of (possibly) different modes.
In a continuum picture, the disconnections are continuously distributed along the GB with density function $\rho_i(x,t)$; this produces a GB shape profile $h_x = \textstyle\sum_i \rho_i H_i$.
Without loss of generality, the density function $\rho_i$ can be positive or negative to represent the density of disconnections of type $(\bm{b}_i, H_i)$ or $(-\bm{b}_i, -H_i)$, respectively.
This may be interpreted to imply that a positive $(\bm{b}_i, H_i)$ and a negative  $(-\bm{b}_i, -H_i)$ disconnection at the same location will automatically annihilate, while disconnections of different modes do not react, in general, along the GB.

Disconnections may nucleate in pairs (conservation of disconnection character) and glide within the GB plane (we do not consider disconnection climb here).
Assuming that disconnection glide is  overdamped, the disconnection glide velocity 
may be described as
\begin{equation}\label{eq:vd}
v_d = M_d\big[(\sigma_d+\tau)b + (\Psi - \gamma h_{xx})H\big],
\end{equation}
where $M_d$ is the disconnection mobility and the terms in the square bracket represent the total driving force on the disconnection.
The term that multiplies  the Burgers vector ($b$) represents the Peach-Koehler force \cite{anderson2017theory}  on the disconnection, where $\sigma_d$ and $\tau$ are the components of the shear stress  parallel to the disconnection Burgers vector associated with the other disconnections and the applied stress, respectively.
The  stress  $\sigma_d$ is nonlocal and is a function of the GB shape profile and the overall disconnection distribution along the GB.
The terms that multiplies the step height ($H$) include jumps in the chemical potential (bulk energy density, $\Phi$) across the GB, denoted $\Psi$, and the capillary force where $\gamma$ is the GB energy density and $h_{xx}$ accounts for the GB curvature.
(Note that the jump in chemical potential may be written to include the capillary force as per the Gibbs-Thomson relation, but is separated here to explicitly emphasize capillarity effects.)
For simplicity, we consider the GB energy $\gamma$ to be inclination-independent in the remainder of this paper.
These driving forces are simply the variation of the total energy of the system with respect to disconnection displacement.
We demonstrate this in Sec.~\ref{sec:TJ_model}.

Disconnection $(\bm{b},H)$ glide (in the $x$-direction)  translates the GB in its wake by $-H$ (in the $y$-direction) and translates the upper grain relative to the lower grain by $b$. Based on this description, GB migration is associated with the disconnection (step) flux along the GB $h_t=-\sum_i \mathcal{J}_iH_i$, where $\mathcal{J}_i$ is the flux of disconnections of mode $i$.
Counting both positive and negative disconnections of mode $i$ in the flux $\mathcal{J}_i$ yields
\begin{equation}\label{eq:J_i}
\mathcal{J}_i = v_i \big(|\rho_i|+2c_i\big),
\end{equation}
where $v_i$ is the glide velocity of disconnection mode $i$.
The term $c_i$ accounts for the thermal equilibrium concentration of disconnections of mode $i$ \cite{zhang2017prl},
\begin{equation}\label{eq:c_i}
c_i = \frac{1}{a} e^{-E_i/(k_B T)},
\end{equation}
where $a$ is an atomic spacing, $E_i$ is half  the disconnection pair formation energy,  $k_B$ is the Boltzmann constant and $T$ is temperature.
The  formation energy of a disconnection pair depends on both its Burgers vector and step height; this  can be estimated  analytically \cite{han2018} or determined by fitting to atomistic simulation results \cite{chen2019shear}.
Using the glide velocity Eq.~(\ref{eq:vd}) and the disconnection flux Eq.~(\ref{eq:J_i}), we can now write the equation of motion for the GB as
\begin{equation}\label{eq:v_gb1}
h_t = -\sum_i M_i \big[(\sigma_d+\tau)b_i + (\Psi - \gamma h_{xx})H_i\big]\big(|\rho_i|+2c_i\big)H_i,
\end{equation}
where $M_i$ is the  mobility of a disconnection of mode $i$, which is, in general, different for different disconnection modes (with different temperature dependences).
For simplicity and illustrative purposes,  we assume here that all disconnection modes have identical, constant mobilities $M_i=M_d$ and do not explicitly consider the temperature dependence of the disconnection mobility while focussing on the temperature dependence of the equilibrium disconnection density $c_i$.
The evolution of the densities of disconnections of each mode along the GB is described by a continuity equation that conserves Burgers vector and step height
\begin{equation}\label{eq:evo_dis}
\frac{\partial \rho_i}{\partial t}=-\frac{\partial}{\partial x}\big[v_i(|\rho_i|+2c_i)\big], \quad \text{for $i=1,2, \ldots$.}
\end{equation}

GB migration, in this disconnection mechanism-respecting model,  is described in terms of the fluxes and the thermal equilibrium concentration of disconnections.
The importance of introducing a disconnection-based description of GB migration is twofold.
First,  stress-driven GB migration is associated with the coupling between disconnection Burgers vectors and  stress; such effects cannot be captured or explained by  conventional capillarity-driven GB migration models.
Second, because the disconnection model accounts for bicrystallography (allowed sets of Burgers vectors and step heights), this approach intrinsically describes the different inherent dependence of the dynamics of different GBs based on the macroscopic degrees of freedom of a GB (misorientation, inclination)  and temperature.
In such a model, the ``apparent" GB mobility is traced to the underlying disconnection properties and the competition between disconnections of different modes (which varies depending on LOCAL driving forces).
This disconnection model for GB migration inspires a novel description of TJ dynamics in terms of the thermal equilibrium concentration  and migration of disconnections.

\section{Disconnection model for triple junction motion}
\label{sec:TJ_model}
In this section, we develop a disconnection-based description of triple junction dynamics.
Consider a system of three grains $\Omega_I$ and three grain boundaries GB$^{(j)}$ meeting at a triple junction (see Fig.~\ref{fig:TJ_fig}).
We describe the profile of GB$^{(j)}$ as $\textbf{X}^{(j)}=(x^{(j)}(s,t), y^{(j)}(s,t))$ in 2D,  where the superscript $j$ denotes different GBs and the parameter $0\leq s\leq 1$ describes positions along the GBs such that $s=0$ and $1$ corresponds to the two TJs that delimit the GB.
GB$^{(j)}$ has multiple possible disconnection modes $(\bm{b}^{(j)}_i, H^{(j)}_i)$ for $1\leq i \leq N_j$ (in principle, $N_j=\infty$)
, where $\bm{b}^{(j)}_i = b^{(j)}_i \bm{e}^{(j)}$ is parallel to the reference GB direction $\bm{e}^{(j)}$ and $H^{(j)}_i$ is the step height (along the reference direction normal, $\bm{n}^{(j)}$).
Since the signed disconnection density function  represents both positive and negative disconnections, we associate  $b^{(j)}_i>0$ with $\rho^{(j)}_i>0$ (without loss of generality).

We use a variational approach, based on the principle of maximum energy dissipation, to derive the equations of  GB/TJ motion.
The GB/TJ network evolves to reduce the total energy in a steepest descent (overdamped) sense; the total energy includes the local energy of the GB/TJ system, the long-range elastic interaction energy of disconnections along the GBs, the energy associated with the TJs, and the (local) elastic energy associated with the applied stress $\bm{\tau}$, i.e., $E_{tot}= E_{local}+E_{long}+E_{TJ}+E_{\tau}$.
The local energy consists of the GB energy and the grain bulk energy
\begin{equation}\label{eq:E_local}
E_{local} = \sum_{j=1}^3 \int_0^1 \gamma^{(j)} \lVert \bm{l}^{(j)} \rVert ds + \sum_{I=1}^3 \iint_{\Omega_I}\Phi_I dA,
\end{equation}
where $\bm{l}^{(j)}=\partial \textbf{X}^{(j)}/\partial s$, $\Phi_I$ is the bulk energy of grain $I$.
The long-range elastic interactions between all disconnections along the GBs is \cite{anderson2017theory}
\begin{equation}\label{eq:E_long}
E_{long} =\! \frac{K}{2}\! \sum_{j=1}^3\sum_{k=1}^3\!\int_0^1\!\!ds^{(j)}\!\!\int_0^1\!\!ds^{(k)}\! \Big(\!\sum_i\rho^{(j)}_i\!\bm{b}^{(j)}_i\!\times\bm{\xi}\Big) \cdot \Big(\!\nabla\otimes\nabla r_{jk}\Big) \cdot
\Big(\!\sum_i\!\rho^{(k)}_i\!\bm{b}^{(k)}_i\!\times\bm{\xi}\Big),
\end{equation}
where $K\equiv \mu/[4\pi(1-\nu)]$ ($\mu$ is the shear modulus and $\nu$ is the Poisson ratio) and $\bm{\xi}$ is a unit vector in the direction of the disconnection line (perpendicular to the two dimensional GB profile).
Here $r_{jk}=\|\textbf{X}^{(j)}-\textbf{X}^{(k)}\|$ and the tensor $\nabla\otimes\nabla r_{jk}$ is taken with respect to $\textbf{X}^{(j)}$.
The prefactor  ($1/2$) corrects for double counting  the  interactions between  disconnections.
If there is an accumulated Burgers vector $\bm{b}_{TJ}$ at the TJ, the associated energy is
\begin{equation}\label{eq:E_TJ}
E_{TJ} = C \|\bm{b}_{TJ}\|^2
+K \sum_{j=1}^3 \int_0^1
\Big(\sum_i\rho^{(j)}_i\bm{b}^{(j)}_i\times\bm{\xi}\Big) \cdot
\Big(\nabla\otimes\nabla r_{j,TJ}\Big) \cdot
\Big(\bm{b}_{TJ}\times\bm{\xi}\Big) ds,
\end{equation}
where $C>0$ accounts for the TJ core energy and is assumed to be independent of disconnection mode here.
The total rate of change of the energy of the system associated with the motion of the TJ (at $s=0$) and its GBs is
\begin{equation}
\begin{aligned}
\dot{E}_{tot} =
& -\sum_{j=1}^3 \int_0^1  (\gamma^{(j)} \kappa^{(j)}-\Psi^{(j)})\lVert \bm{l}^{(j)} \rVert \bm{\eta}^{(j)} \cdot \bm{v}^{(j)}_{GB} ds - \sum_{j=1}^3 \gamma^{(j)}\bm{t}^{(j)}(0)\cdot\bm{v}_{TJ} \\
&-\sum_{j=1}^3 \sum_i \int_0^1\! |\rho^{(j)}|\big[(\bm{\sigma}^{(j)}\cdot\bm{b}^{(j)}_i)\times\bm{\xi}\big]\cdot \bm{v}^{(j)}_{i} ds  \\
& + 2C\bm{b}_{TJ}\cdot \dot{\bm{b}}_{TJ} - \big[(\bm{\sigma}_{TJ}\cdot\bm{b}_{TJ})\times\bm{\xi}\big]\cdot\bm{v}_{TJ},
\label{eq:E_total_dot}
\end{aligned}
\end{equation}
where $\Psi^{(j)}$ is the jump of bulk energy density across GB$^{(j)}$ (e.g., $\Psi^{(1)}=\Phi_3-\Phi_2$), $\bm{t}^{(j)}=\bm{l}^{(j)}/\|\bm{l}^{(j)}\|$ is the unit tangent vector along the GB$^{(j)}$, $\bm{\sigma}^{(j)}$ and $\bm{\sigma}_{TJ}$ are the total stresses acting at a point along GB$^{(j)}$ and at the TJ
(the total stresses  have non-local contributions from all of the disconnections in the system and the applied stress, 
$\bm{\sigma}=\bm{\sigma}_d+\bm{\tau}$), respectively.
The total rate of change of the energy of the entire GB/TJ network may be found by summing over all TJs.
The first line in Eq.~\ref{eq:E_total_dot}  describes the rate of change of the local energy ($E_{local}$), where we use integration by parts for the GB energy term. 
The second line is associated with the work done by the Peach-Koehler force on the disconnections from the other disconnections along the GBs ($E_{long}$) and at the TJ ($E_{TJ}$), and as well as the applied stress ($E_{\tau}$).
The last line corresponds to the rate of change of the energies associated with the accumulated Burgers vector at the TJ  ($E_{TJ}$ and $E_{\tau}$).

Assuming, as in Sec.~\ref{sec:GB_model}, that the GB only migrates in the direction normal to the reference GB  (disconnection glide parallel to the reference GB), the glide of  positive disconnections $\bm{v}^{(j)}_{i}=v^{(j)}_{i}\bm{e}^{(j)}$ produces GB migration with velocity $\bm{v}_{GB}^{(j)}= - \sum_i v^{(j)}_{i}|\rho^{(j)}_{i}|H^{(j)}_{i}\bm{n}^{(j)}$.
We can then rewrite Eq.~(\ref{eq:E_total_dot}) as
\begin{equation}\label{eq:Etot_der}
\begin{aligned}
\dot{E}_{tot} =
& -\sum_{j=1}^3 \sum_i \int_0^1 \Big[(\bm{\sigma}^{(j)}\cdot\bm{b}^{(j)}_i\times\bm{\xi})\cdot \bm{e}^{(j)}- (\gamma^{(j)} \kappa^{(j)}-\Psi^{(j)})H^{(j)}_{i} \Big] v^{(j)}_{i} |\rho_i^{(j)}| ds  \\
& - \Big[\sum_{j=1}^3 \gamma^{(j)}\bm{t}^{(j)}+(\bm{\sigma}_{TJ}\cdot\bm{b}_{TJ})\times\bm{\xi}\Big]\cdot\bm{v}_{TJ} + 2C\bm{b}_{TJ}\cdot \dot{\bm{b}}_{TJ},
\end{aligned}
\end{equation}
where we applied the fact $\lVert \bm{l}^{(j)} \rVert \bm{\eta}^{(j)}\cdot\bm{n}^{(j)}=1$ to the second term in the first integral.

We introduce the dissipation function associated with the TJ and disconnection motion:
\begin{equation}
\label{eq:Q_dissipation}
Q = \sum_{j=1}^3 \sum_i \int_0^1 \frac{(v^{(j)}_{i})^2}{M_d} |\rho^{(j)}_i| ds + \frac{\|\bm{v}_{TJ}\|^2}{M_{TJ}} + \frac{\|\dot{\bm{b}}_{TJ}\|^2}{\tilde{C}},
\end{equation}
where $\tilde{C}$ is a coefficient associated with the energy dissipation due to the change of accumulated Burgers vector at TJ; $\tilde{C}$ can be thought of as related to a kinetic  barrier for disconnections entering (or leaving) the TJ or reacting at the TJ.
By the maximum dissipation  principle, we describe the evolution of the TJ/GB network as that which maximizes the energy dissipation $Q$ subject to the balance condition
\begin{equation}\label{eq:dissipation_balance}
Q+\dot{E}_{tot}=0.
\end{equation}

In addition to the balance condition, we have two compatibility conditions for the disconnection flux between the TJ and its constituent GBs (Fig.~\ref{fig:TJ_velocity}) that account for the constraints  of  GB/TJ crystallography on  TJ motion.
To guarantee that the three constituent GBs remain connected to their TJ during the evolution of GB/TJ system, the TJ motion must be geometrically consistent with the migration of the adjacent GBs (written in terms of the step flux):
\begin{equation}\label{eq:step_comp}
\sum_{i} J_i^{(j)}H_i^{(j)}
= \bm{v}_{TJ}\cdot\bm{n}^{(j)} \qquad \text{for $j=1,2,3$},
\end{equation}
where $J_i^{(j)}$ represents the flux of disconnections of mode $i$ from GB$^{(j)}$ into the TJ.
Meanwhile, the Burgers vector carried by the disconnection flux into the TJ consequently causes Burgers vector accumulation at the TJ:
\begin{equation}\label{eq:b_comp}
\dot{\bm{b}}_{TJ}
= \sum_{j=1}^{3} \sum_{i} J_i^{(j)}\bm{b}_i^{(j)}.
\end{equation}

We can now construct the Lagrangian for this constrained optimization problem
\begin{equation}
L_v = Q + \lambda \Big(Q + \dot{E}_{tot}\Big)
+ \alpha \Big\|\dot{\bm{b}}_{TJ} - \sum_{j=1}^{3}\sum_{i} J_i^{(j)}\bm{b}_i^{(j)}\Big\|^2
+ \sum_{j=1}^{3} \beta_j \Big(\sum_{i} J_i^{(j)}H_i^{(j)}-\bm{v}_{TJ}\cdot\bm{n}^{(j)}\Big),
\end{equation}
where $\lambda$, $\alpha$, $\beta_j$ are  Lagrange multipliers, and $v^{(j)}_{i}$, $\bm{v}_{TJ}$, $\dot{\bm{b}}_{TJ}$ and $J_i^{(j)}$ are the kinetic variables. The stationary equations are
\begin{align}
\label{eq:L_vij}
&\frac{\delta L_v}{\delta v^{(j)}_{i}}
= -\lambda  \Big[(\bm{\sigma}^{(j)}\cdot\bm{b}^{(j)}_i\times\bm{\xi})\cdot \bm{e}^{(j)}-(\gamma^{(j)}\! \kappa^{(j)}\!-\!\Psi^{(j)})H^{(j)}_{i}\!\Big]
+ 2(1+\lambda) \frac{v^{(j)}_{i}}{M_d} = 0,
\\
\label{eq:L_vtj}
&\frac{\delta L_v}{\delta \bm{v}_{TJ}}
= -\lambda \Big[\sum_j \gamma^{(j)}\bm{t}^{(j)}\!+\!(\bm{\sigma}_{TJ}\cdot\bm{b}_{TJ})\times\bm{\xi}\Big]
 + 2(1+\lambda)\frac{\bm{v}_{TJ}}{M_{TJ}} - \sum_{j=1}^{3} \beta_j\bm{n}^{(j)} =\bm{0},
\\
\label{eq:L_b}
&\frac{\delta L_v}{\delta \dot{\bm{b}}_{TJ}}
= 2\lambda C\bm{b}_{TJ}+2(1+\lambda)\frac{\dot{\bm{b}}_{TJ}}{\tilde{C}} +2\alpha \Big(\dot{\bm{b}}_{TJ}-\sum_{j=1}^{3} \sum_{i} J_i^{(j)}\bm{b}_i^{(j)}\Big)=\bm{0},
\\
\label{eq:L_Jij}
&\frac{\delta L_v}{\delta J_i^{(j)}}
= -2\alpha\Big(\dot{\bm{b}}_{TJ}-\sum_{j=1}^{3} \sum_{i} J_i^{(j)}\bm{b}_i^{(j)}\Big)\cdot\bm{b}_i^{(j)} + \beta_jH_i^{(j)}=0.
\end{align}
Inserting the Burgers vector flux compatibility condition (Eq.~\ref{eq:b_comp}) into Eq.~(\ref{eq:L_Jij}) yields $\beta_j=0$ for $j=1,2,3$.
By comparing Eq.~\ref{eq:dissipation_balance} (after inserting terms from Eqs.~\ref{eq:Etot_der} and \ref{eq:Q_dissipation}) with $2(1+\lambda)Q+\lambda\dot{E}_{tot}=0$, 
shows that $\lambda=-2$.
The expression $2(1+\lambda)Q+\lambda\dot{E}_{tot}=0$ was found from the addition of three terms:  ({\it{i}}) multiply Eq.~(\ref{eq:L_vij}) by $v^{(j)}_i|\rho^{(j)}_i|$, then sum over $i$ and $j$, and integrate along the GB, ({\it{ii}}) multiply Eq.~(\ref{eq:L_vtj}) by $\bm{v}_{TJ}$, and ({\it{iii}}) multiply Eq.~(\ref{eq:L_b}) by $\dot{\bm{b}}_{TJ}$.

Combining these results, we obtain the equations governing the evolution of the system (disconnection glide velocities, TJ velocity, and Burgers vector accumulation rate at the TJ):
\begin{align}
\label{eq:vij}
v^{(j)}_{i} &= M_d\Big[(\bm{\sigma}^{(j)}\cdot\bm{b}^{(j)}_i\times\bm{\xi})\cdot \bm{e}^{(j)}-(\gamma^{(j)} \kappa^{(j)}-\Psi^{(j)})H^{(j)}_{i} \Big] ,\\
\label{eq:vtj}
\bm{v}_{TJ} &= M_{TJ}\Big[\sum_j \gamma^{(j)}\bm{t}^{(j)}+(\bm{\sigma}_{TJ}\cdot\bm{b}_{TJ})\times\bm{\xi}\Big], \\
\label{eq:dBtj}
\dot{\bm{b}}_{TJ} &= -2\tilde{C}C\bm{b}_{TJ}.
\end{align}

Equation~(\ref{eq:vij}) describes GB dynamics.  The disconnection glide velocity in Eq.~(\ref{eq:vd}) is recovered from Eq.~(\ref{eq:vij}) with the small-slope approximation for the GB inclination (recall that  the total stress $\bm{\sigma}^{(j)}=\bm{\sigma}_d^{(j)}+\bm{\tau}$).
The equation describing GB migration (Eq.~\ref{eq:v_gb1}) is obtained by inserting Eq.~(\ref{eq:vij}) into the disconnection flux and accounting for the equilibrium disconnection density.

When there is no Burgers vector accumulation at the TJ ($\bm{b}_{TJ}=\bm{0}$), $\dot{\bm{b}}_{TJ} =0$ (see Eq.~\ref{eq:dBtj}) and Eq.~(\ref{eq:vtj}) reduces to the classical expression for TJ motion in the capillarity-driven model (isotropic GB energy case).
When the Burgers vector at the TJ is non-zero, Eq.~(\ref{eq:dBtj}) shows that the system evolves  to eliminate the  Burgers vector at the TJ  ($\tilde{C}C\geq0$) and hence relax the stress field associated with TJ; this is consistent with  observations from  molecular dynamics simulations  \cite{thomas2019TJ}.
Comparing  Eqs.~(\ref{eq:vtj}) and (\ref{eq:dBtj}) suggests that  $\tilde{C}$ is associated with the rate of relaxation of the  Burgers vector at the TJ and $-2C\bm{b}_{TJ}$ is the thermodynamic driving force for that relaxation.
Note that  jumps in the chemical potential $\Psi^{(j)}$ do not appear in the first variation of the energy with respect to TJ displacement, and hence does not affect the TJ dynamics (i.e., it contributes to higher order variations).

Equations~\ref{eq:vij}-\ref{eq:dBtj}, describing the macroscopic evolution of the TJ/GB network, have some interesting implications for TJ motion at the microscopic level.
While TJ motion is dictated by the disconnection flux between the TJ and its constituent GBs, there are implicit constraints on the flux of disconnections as seen by comparing Eqs.~(\ref{eq:vtj})-(\ref{eq:dBtj}) with  compatibility conditions Eqs.~(\ref{eq:step_comp})-(\ref{eq:b_comp}):
\begin{align}
\label{eq:step_constraint}
&\sum_{i=1}^{N_j} J_i^{(j)}H_i^{(j)} = M_{TJ} \Big[\sum_k \gamma^{(k)}\bm{t}^{(k)}+(\bm{\sigma}_{TJ}\cdot\bm{b}_{TJ})\times\bm{\xi}\Big]\cdot\bm{n}^{(j)}, \quad j=1,2,3,
\\
\label{eq:b_constraint}
&\sum_{j=1}^{3} \sum_{i=1}^{N_j} J_i^{(j)}\bm{b}_i^{(j)} = -2\tilde{C}C\bm{b}_{TJ}.
\end{align}
These constraints represent five equations (in two dimensions) for a set of $\textstyle\sum_{j=1}^3 N_j$ flux variables $J_i^{(j)}$ (recall that $N_j$ is the number of disconnection modes active on GB $j$).
If there is only one disconnection mode operating on each of the constituent GBs ($N_j=1$), the linear system of constraint conditions  is overdetermined.
This implies that for general GBs meeting at TJs there are no single disconnection mode solutions and hence the TJ/GB will not evolve (except for a small set of very special cases).
On the other hand, if multiple disconnection modes are active on each GB ($N_j\geq 2$), the system is underdetermined and evolution is possible (i.e., the disconnection fluxes are not determined solely by these constraints).
This implies that multiple disconnection modes are required for polycrystalline microstructure evolution.

Since disconnections must form before they can migrate, disconnection fluxes are further limited by the rate at which disconnection are formed (i.e., the space of the flux variables is bounded). This implies that the  disconnection flux into the TJ is limited as
\begin{equation}\label{eq:flux_limit}
|J_i^{(j)}| \leq   c_i^{(j)} M_d \Big|2C\bm{b}_{TJ}\cdot\bm{e}^{(j)}
+\sgn(H_i^{(j)})\Big(\sum_k \gamma^{(k)}\bm{t}^{(k)}+\bm{\sigma}_{TJ}\cdot\bm{b}_{TJ}\times\bm{\xi}\Big)\cdot\bm{n}^{(j)}\Big|,
\end{equation}
where the term in the absolute value is the force on the disconnections at the triple junction.

This condition describes the maximum possible disconnection flux that can contribute to the TJ motion.
The limitation on the flux in Eq.~(\ref{eq:flux_limit}) depends on the availability of disconnections (limited by  $c_i^{(j)}$), the disconnection mobility $M_d$, and the disconnection step heights and Burgers vectors.
Note that the absolute value in this equation accounts for both positive and negative disconnections and the signum function $\sgn(x)$ accounts for both positive and negative step heights.
This limitation condition on the disconnection flux complicates consideration of the existence and uniqueness of  flux solutions that satisfy the constraints.
Even for the case of multiple disconnection modes, this limitation implies that it is possible that no solutions exist (i.e., the disconnection flux required by the constraints (Eqs.~\ref{eq:step_constraint}--\ref{eq:b_constraint}) may exceed the limitation of the disconnection flux).

This analysis suggests possible inconsistency between the macroscopic equations for triple junction dynamics and the limitations on the microscopic disconnection flux.
This observation suggests the need to reconsider the notion of triple junction mobility.
Just as the GB migration is not a monolithic thermally-activated process but rather an aggregate of thermally-activated disconnection formation and migration events, the same should be true for the triple junction.
Therefore, the TJ mobility is not an intrinsic property of a TJ, but depends on the disconnection properties and their responses to the local, spatio-temporal environment.
In other words, the TJ mobility can not only vary between TJs due to different GB/TJ crystallography but also vary   during TJ migration as a result of changes in the environment (e.g., stress accumulation or relaxation, the GB/TJ local structure).
Hence, the TJ mobility $M_{TJ}$ should not be viewed as an \textit{a priori} prescribed constant.
A similar argument can  also be applied to the parameter $\tilde{C}$ describing TJ Burgers vector relaxation.
If we describe  $M_{TJ}$ and $\tilde{C}$ as variable parameters (to be determined later), the two constraints on the disconnection flux Eqs.~(\ref{eq:step_constraint})--(\ref{eq:b_constraint}) are relaxed.
While the space of flux variables is bounded due to the limitation on disconnection flux, the two constraint conditions can always be satisfied within the bounded space by varying these parameters.
To determine the mobility parameters $M_{TJ}$ and $\tilde{C}$ (which are now functions of the disconnection flux), we recall that the system evolves in a manner that maximizes  energy dissipation.

Now determination of the  TJ dynamics becomes an optimization problem of maximizing the energy dissipation function with respect to the disconnection flux at the TJ, $J_i^{(j)}$ (where $M_{TJ}\geq 0$ and $\tilde{C}\geq 0$ are dependent on the  disconnection flux):
\begin{equation} \label{eq:obj}
\max_{J^{(j)}_i}\tilde{Q} = M_{TJ}
\Big\| \sum_{k} \gamma^{(k)} \bm{t}^{(k)}+\bm{\sigma}_{TJ}\cdot\bm{b}_{TJ}\times\bm{\xi}\Big\|^2 +\ \tilde{C}\Big\| 2C\bm{b}_{TJ}\Big\|^2
\end{equation}
subject to   constraints
\begin{align} \label{eq:constr1}
&\sum_{j=1}^{3} \sum_{i=1}^{N_j} J_i^{(j)}\bm{b}_i^{(j)} = -2\tilde{C}C\bm{b}_{TJ},
\\
\label{eq:constr2}
&\sum_{i=1}^{N_j} J_i^{(j)}H_i^{(j)} = M_{TJ} \Big[\sum_k \gamma^{(k)}\bm{t}^{(k)}+(\bm{\sigma}_{TJ}\cdot\bm{b}_{TJ})\times\bm{\xi}\Big]\cdot\bm{n}^{(j)},  \text{ for $j=1,2,3,$} \\
\label{eq:constr3}
&|J_i^{(j)}| \! \leq \! M_d  c_i^{(j)} \Big|2C\bm{b}_{TJ}\cdot\bm{e}^{(j)} \!
+\sgn(H_i^{(j)})\Big(\!\sum_k \gamma^{(k)}\bm{t}^{(k)}+\bm{\sigma}_{TJ}\cdot\bm{b}_{TJ}\times\bm{\xi}\Big)\cdot\bm{n}^{(j)}\Big|.
\end{align}
This is a linear optimization problem and can be  solved by, for example, linear programming methods.
The resultant TJ dynamics is described by the governing equations with $M_{TJ}$ and $\tilde{C}$ determined from the optimization problem
\begin{equation}\label{eq:vtj2}
\bm{v}_{TJ} = M_{TJ}\Big[\sum_j \gamma^{(j)}\bm{t}^{(j)}+(\bm{\sigma}_{TJ}\cdot\bm{b}_{TJ})\times\bm{\xi}\Big], \qquad
\dot{\bm{b}}_{TJ} = -2\tilde{C}C\bm{b}_{TJ}.
\end{equation}
The resultant GB (and disconnection) dynamics is then governed by
\begin{align}\label{eq:GB_evol}
h_t^{(j)}(s) = -\sum_i \mathcal{J}_i^{(j)}(s)H_i^{(j)},\qquad
\frac{\partial }{\partial t}\rho_i^{(j)} = -\frac{\partial}{\partial s}\mathcal{J}_i^{(j)}(s), \quad \text{for $0<s<1$,}\\
\label{eq:flux}
\mathcal{J}_i^{(j)}(s) = M_d \Big[(\bm{\sigma}^{(j)}\cdot\bm{b}^{(j)}_i\times\bm{\xi})\cdot \bm{e}^{(j)} -(\gamma^{(j)} \kappa^{(j)}-\Psi^{(j)})H^{(j)}_{i} \Big]\big(|\rho_i^{(j)}|+2c_i^{(j)}\big),
\end{align}
subject to the boundary conditions at the TJ ($s=0$)
\begin{equation} \label{eq:bc}
h_t^{(j)}(s=0) = \bm{v}_{TJ}\cdot\bm{n}^{(j)},  \quad  \mathcal{J}_i^{(j)}(s=0) = -J_i^{(j)},
\end{equation}
where $J_i^{(j)}$ is the solution of the optimization problem.
The boundary condition at the far end of the GBs ($s=1$) will be similar but associated with other TJs.

\section{Analysis and simulation}\label{sec:simulation}
\subsection{Burgers vector accumulation and relaxation}
\label{subsec:B_accumulation}
In the disconnection model of TJ dynamics, we consider  the effects of the accumulated Burgers vector on triple junction motion.
The presence of a non-zero Burgers vector at a TJ introduces a Peach-Koehler force on the TJ that, in addition to capillary forces, dictates the TJ trajectory as described in Eq.~(\ref{eq:vtj}); the TJ migrates in a direction that reduces the total GB energy and elastic  energy of the system.
The dynamics of Burgers vector accumulation and relaxation described by Eq.~(\ref{eq:dBtj}) shows that each TJ is driven to zero TJ Burgers vector.
Any nonzero TJ  Burgers vector will be relaxed via reaction with the disconnection flux from the GBs.
Molecular dynamics simulation results demonstrate that the relaxation of the Burgers vector at a TJ occurs very quickly compared with GB/TJ migration \cite{thomas2019TJ}.
Therefore, it is reasonable to employ a zero Burgers vector  condition at the TJ in our continuum model (in most situations).

Burgers vector accumulation/relaxation has an important  influence on  triple junction mobility.
In order to maintain the zero Burgers vector  condition, multiple disconnection modes along each GB are necessary for TJ migration; otherwise, TJ migration stagnates.
If there is only one disconnection mode $N_j=1$, the admissible set of solutions that satisfy  constraints Eqs.~(\ref{eq:constr1})--(\ref{eq:constr3}) is empty when $M_{TJ}\neq 0$ and/or $\tilde{C}\neq 0$ (except for some special cases).
When $M_{TJ}= 0$ and $\tilde{C}= 0$, the system of constraint equations is homogeneous and only the zero solution exists, i.e., $J^{(1)}=J^{(2)}=J^{(3)}=0$; this indicates  stagnation of the TJ dynamics (i.e., zero TJ mobility and zero disconnection  exchange between GBs and the TJ).
To \mbox{fulfill} the constraints during TJ migration, we require disconnections of different modes ($N_j\geq 2$) in order to relax the accumulated Burgers vector at the TJ; there may exist nonzero solutions for nonzero $M_{TJ}$ and $\tilde{C}$ by adding degrees of freedoms (other disconnection modes) in the linear system.
Even with multiple disconnection modes, however, the TJ mobility may be very small; it strongly depends on the relative ease of nucleating disconnections of additional modes which determines the size of the bounded space of flux variables in Eq.~(\ref{eq:constr3}).
We  return to this issue in Section \ref{subsec:M_tj}.

Another possible mechanism by which TJs may relax an accumulated \mbox{Burgers} vector is to emit defects into the surrounding grains such as lattice dislocations \cite{Hashimoto1987,Chen2003science,Liao2003,VanSwygenhoven2002dislocationemission} or  twins \cite{VanSwygenhoven2001twin,Mason2015geometric,thomas2016twin,lin2015}.
Suppose that a TJ is able to emit dislocations in $M$ slip systems into the surrounding grains; the  Burgers vector relaxation dynamics Eq.~\ref{eq:b_constraint} becomes
\begin{equation} \label{eq:emit}
\sum_{j=1}^{3} J^{(j)}\bm{b}^{(j)} + \sum_{m=1}^{M} J^{m}\bm{b}^{m}= -2\tilde{C}C\bm{b}_{TJ},
\end{equation}
where $J^m$ and $\bm{b}^m$ are the flux and Burgers vector of the lattice dislocation emitted on slip system $m$.
Given that the TJ can only emit dislocations on a limited set of slip planes, this mechanism of Burgers vector relaxation via dislocation emission is only viable when the slip systems are suitably oriented.

\subsection{Triple junction mobility}
\label{subsec:M_tj}
We consider an example of a triple junction with three initially flat GBs in a tri-grain as shown in Fig.~\ref{fig:TJ_circle}.
We first examine a relatively simple case in which the reference inclination of GB$^{(1)}$ is vertical and the crystallography of GB$^{(2)}$ and GB$^{(3)}$ are mirror-symmetric in terms of GB reference inclination ($\phi^{(2)}=\phi^{(3)}$ as seem in Fig.~\ref{fig:TJ_angles}) and disconnection properties. We further assume that initially there is no Burgers vector at the TJ ($\bm{b}_{TJ}=\bm{0}$), the GB/TJ profile is symmetric such that GB$^{(1)}$ is along its reference inclination and the GB$^{(2)}$ and GB$^{(3)}$ inclinations  deviate only slightly from their reference inclinations and lie symmetrically about GB$^{(1)}$ ($\theta^{(2)}=\theta^{(3)}$).
We consider two disconnection modes on each GB, with different coupling factors, i.e., $b_1^{(j)}/H_1^{(j)}\neq b_2^{(j)}/H_2^{(j)}$.
The assumed symmetries enable us to obtain an analytical expression for the TJ mobility.
There should be no net disconnection flux from GB$^{(1)}$ because $(\sum_k \gamma^{(k)}\bm{t}^{(k)})\cdot\bm{n}^{(1)}=0$ in Eq.~(\ref{eq:constr3}) and, hence, the disconnection flux from GB$^{(2)}$ (or GB$^{(3)}$) should carry zero net Burgers vector into the TJ.
With these assumptions we find:
\begin{align}
&\bm{v}_{TJ}
= M_{TJ} \sum_{j=1}^3 \gamma^{(j)} \bm{t}^{(j)}, \\
\label{eq:M_tj_two}
&M_{TJ} = \left\{\begin{array}{ll}
M_d c_2
\left|H_2\left(\frac{b_2H_1}{b_1H_2}-1\right)\right|, & \text{if $c_1b_1\geq c_2b_2$} \\[10pt]
M_d c_1
\left|H_1\left(1-\frac{b_1H_2}{b_2H_1}\right)\right|, & \text{if $c_1b_1<c_2b_2$},
\end{array}\right.
\end{align}
where the Burgers vectors and step heights for GB$^{(2)}$ and GB$^{(3)}$ are identical (and written without superscripts above).

While this analytical result is obtained for a symmetric GB/TJ profile with two disconnection modes in each mobile GB, it clearly demonstrates the TJ mobility is a function of the underlying disconnection kinetic properties ($M_d$,  $c_i^{(j)}$), and those disconnection properties are determined by the crystallography $(\bm{b}_i^{(j)},H_i^{(j)})$.
In the single disconnection mode ($c_2^{(2)}=0$) case, the TJ mobility is zero, as expected.
When there are two disconnection modes, the TJ mobility is limited by the availability of a secondary disconnection mode ($b_2^{(j)}/H_2^{(j)}\neq b_1^{(j)}/H_1^{(j)}$).
At low temperature, the equilibrium concentration of secondary  mode disconnections  is expected to be much lower compared with that of the primary disconnection mode ($c_2^{(2)}\ll c_1^{(2)}$); this implies a very small TJ mobility.
At high temperature, on the other hand, the disparity between the nucleation of two modes will be much reduced ($c_2^{(2)}\sim c_1^{(2)}$) such that the TJ mobility is determined by properties of both disconnection modes.
When there are more than two disconnection modes, the effect of higher order modes with even smaller equilibrium concentrations will likely be (relatively) small perturbations to the TJ mobility obtained above on the basis of the primary and secondary disconnections.

For asymmetric TJ profiles, TJ mobility is determined by the disconnections associated with all three constituent GBs. While we cannot analytically determine the TJ mobility for arbitrary, asymmetric TJs (even for two modes); determining such TJ mobilities requires the solution of a multivariable optimization problem.
For the asymmetric TJ profiles ($\bm{b}_{TJ}=\bm{0}$),  constraints Eqs.~(\ref{eq:constr1})-(\ref{eq:constr2}) become
\begin{align} \label{eq:b_balance}
&\Lambda = \frac{\sum_{i} J_i^{(1)}b_i^{(1)}}{\sin\phi^{(1)}}=\frac{\sum_{i} J_i^{(2)}b_i^{(2)}}{\sin\phi^{(2)}}=\frac{\sum_{i} J_i^{(3)}b_i^{(3)}}{\sin\phi^{(3)}},
\\ \label{eq:h_balance}
&v_{TJ}=\frac{\sum_{i} J_i^{(1)}H_i^{(1)}}{\sin\alpha^{(1)}}=\frac{\sum_{i} J_i^{(2)}H_i^{(2)}}{\sin\alpha^{(2)}}=\frac{\sum_{i} J_i^{(3)}H_i^{(3)}}{\sin\alpha^{(3)}},
\end{align}
where $\Lambda$ characterizes the Burgers vector flux into the TJ, $\phi^{(j)}$ is the dihedral angle (between  reference GB planes) opposite GB$^{(j)}$ and $\alpha^{(j)}$ is the angle between $\bm{v}_{TJ}$ and the normal vector $\bm{b}^{(j)}$ (Fig.~\ref{fig:TJ_angles}).
We note that the disconnection fluxes contributing to the TJ mobility in Eq.~(\ref{eq:M_tj_two}) for the symmetric TJ case are such that the total Burgers vector flux from each constituent GB is zero (i.e., $\Lambda=0$ in Eq.~\ref{eq:b_balance}).
Assuming that this also applies in the asymmetric TJ case, we can define a mobility $M_{TJ}^{(j)}$ according to Eq.~(\ref{eq:M_tj_two}) for each constituent GB$^{(j)}$ that satisfies Eq.~(\ref{eq:b_balance}) with $\Lambda=0$ (subject to the disconnection flux limitation Eq.~\ref{eq:constr3}).
These three mobilities, however, do not generally satisfy Eq.~(\ref{eq:h_balance}); they are associated with different values of $v_{TJ}$.
While these TJ mobilities do not satisfy the optimization problem, they provide a lower bound for the TJ mobility
\begin{align}
\label{eq:M_tj_lb}
&M_{TJ} \geq \underline{M}_{TJ}=\min \Big\{M_{TJ}^{(j)}\Big|j=1,2,3\Big\} \\
&M_{TJ}^{(j)} = \left\{\begin{array}{ll}
M_d c_2^{(j)}
\left|H_2^{(j)}\left(\frac{b_2^{(j)}H_1^{(j)}}{b_1^{(j)}H_2^{(j)}}-1\right)\right|, & \text{if $c_1^{(j)}b_1^{(j)}\geq c_2^{(j)}b_2^{(j)}$} \\[10pt]
M_d c_1^{(j)}
\left|H_1^{(j)}\left(1-\frac{b_1^{(j)}H_2^{(j)}}{b_2^{(j)}H_1^{(j)}}\right)\right|, & \text{if $c_1^{(j)}b_1^{(j)}<c_2^{(j)}b_2^{(j)}$}
\end{array}\right..
\end{align}
The resultant TJ mobility $\underline{M}_{TJ}$ is actually the optimal mobility for the special case $\Lambda=0$; the associated flux can be calculated from Eqs.~(\ref{eq:b_balance})--(\ref{eq:h_balance}) and are consistent with the disconnection flux limitation Eq.~(\ref{eq:constr3}).

\subsection{Triple junction drag}
\label{subsec:drag_effect}
When the TJ dynamics is kinetically limited, a triple junction exerts a drag  on  GB migration, as described by the  dimensionless parameter $\delta$ (see Eq.~\ref{eq:drag_delta}), where we now write the characteristic length scale  $w=L_0$ (related to the grain size or distance between TJs along a GB).
Since the present TJ dynamics approach provides a unified description of the GB and TJ mobilities based on the underlying (more fundamental) disconnection properties, it provides a more predictive approach to estimate the TJ drag effect.
While the GB mobility does not explicitly appear in the equation of GB motion Eq.~(\ref{eq:v_gb1}), the ``apparent" GB mobility is the result of the competition between different disconnection modes (under different driving forces).
Recognizing the different driving forces associated with the Burgers vector ($\sigma_d+\tau$) and the step height ($\Psi-\gamma h_{xx}$) in Eq.~(\ref{eq:v_gb1}), we can identify the corresponding GB mobilities. For simplicity and illustrative purpose, we focus this discussion on the GB mobility associated with driving forces that couple to the step height (similar results can be applied to those that couple with the Burgers vector).
In this case, thew GB mobility is
\begin{equation}\label{eq:M_gb}
M_{GB} = M_d \sum_i 2c_i H_i^2,
\end{equation}
where, for simplicity, we  assumed that the net disconnection density of each mode is small ($\rho_i \sim 0$).
Inserting Eqs.~(\ref{eq:M_tj_two}) and (\ref{eq:M_gb}) (two disconnection mode case) into Eq.~(\ref{eq:drag_delta}), we find  the TJ drag  parameter
\begin{equation}\label{eq:drag_delta2}
\delta =
\frac{L_0}{2b_1}
\frac{|b_2H_1-b_1H_2|}{\frac{c_1}{c_2}H_1^2+H_2^2},
\end{equation}
where, without loss of generality, we assume that  $c_1b_1\geq c_2b_2$. At low temperature ($c_1/c_2 \gg 1$), $\delta \rightarrow 0$ ($\delta = 0$ implies that TJs do not move).
This suggests that the TJ drag dominates grain growth at low temperature.
At high temperature, when $c_1/c_2 \sim 1$, $\delta$ is large ($L_0 \gg b_1$). This  implies that  TJ drag is not significant at high temperature and  grain growth is GB migration-controlled; i.e., this is  classical grain growth.

\subsection{Simulation results}
\label{subsec:simulation}
In real materials GBs  have access to multiple disconnection modes.
However, the equilibrium concentration of disconnections varies dramatically for  disconnections of different modes and at different temperatures (see Eq.~\ref{eq:c_i}).
We consider the example of a $\Sigma37$ symmetric tilt GB in Cu.
For this GB, $b_n=na_0/\sqrt{74}$ and $H_{nj}=(6n-37j)a_0/2\sqrt{74}$ \cite{han2018,wei2019gb_mm}, where $a_0$ is the lattice constant, and $n$ and $j$ are integers  that characterize different disconnection modes.
The equilibrium disconnection concentrations are determined based on an analytical expression \cite{han2018} where the parameters were fit to  atomistic simulation data \cite{chen2019shear} using an embedded-atom-method interatomic potential for Cu \cite{Mishin2001Cu};
$2E_i=2E_{nj}= (0.53$ J/m$^2) |H_{nj}| + (36$ GPa) $b_n^2$.
The following parameters were found for Cu and this GB:  shear modulus $\mu=45$ GPa, Poisson's ratio $\nu=0.36$,  lattice constant $a_0=3.615$~\AA\ and GB energy  \cite{cahn2006coupling} $\gamma=0.732$~J/m$^2$.
We  set the GB length scale $L_0=100$~\AA.

\begin{figure}[!ht]
\centering
\subfloat[]{\includegraphics[width=0.33\textwidth]{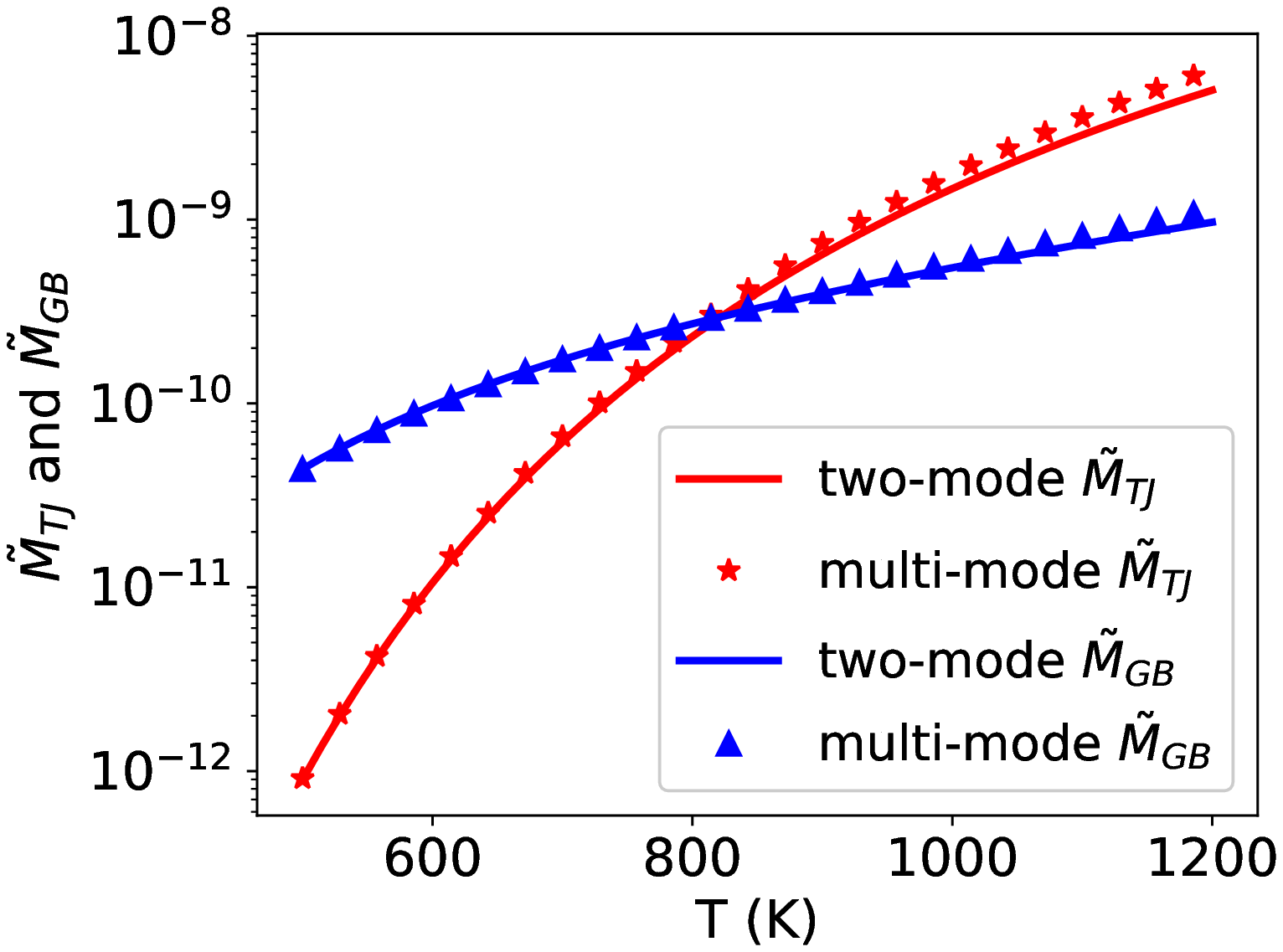}}
\subfloat[]{\includegraphics[width=0.33\textwidth]{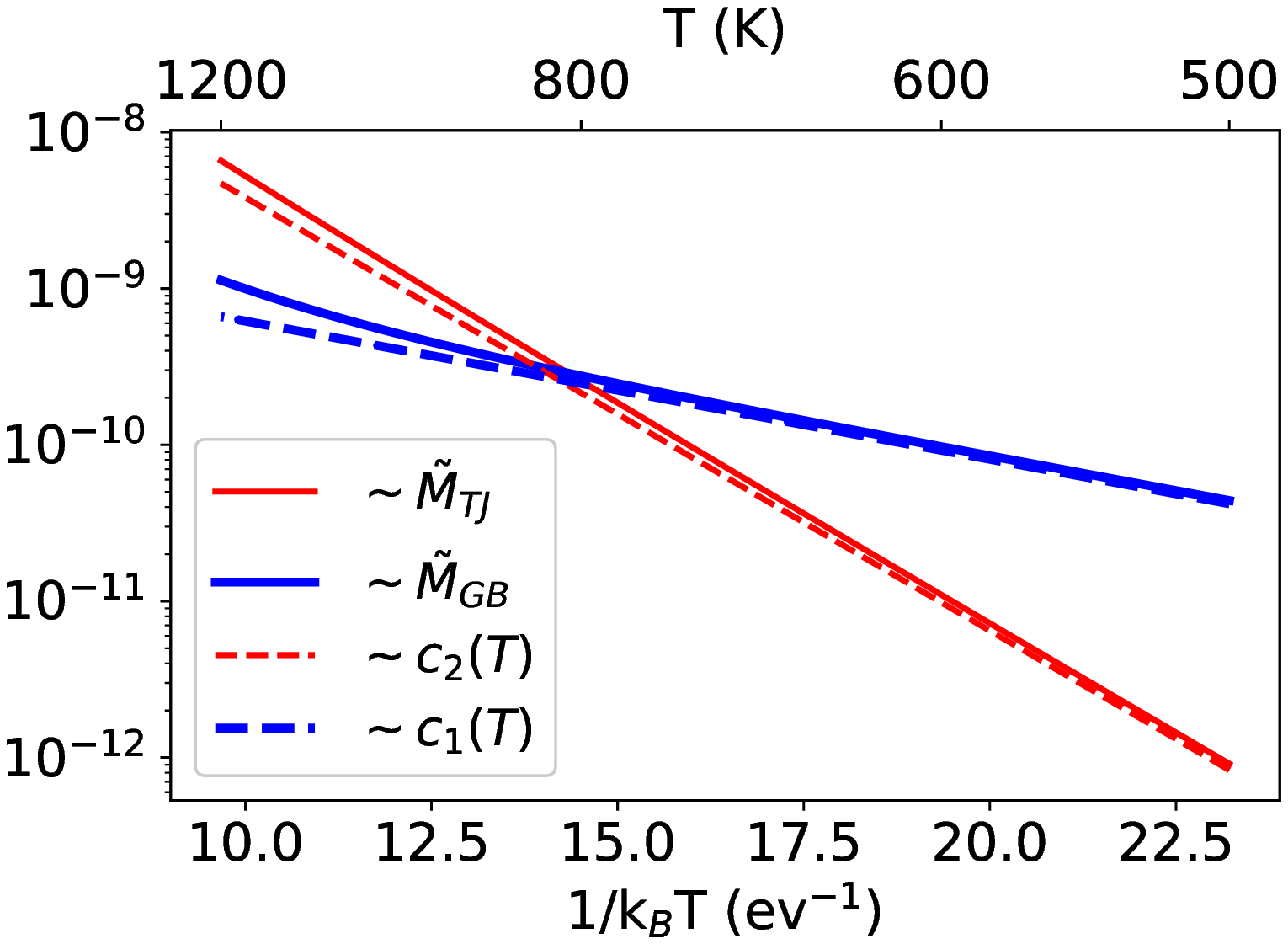}}
\subfloat[]{\includegraphics[width=0.33\textwidth]{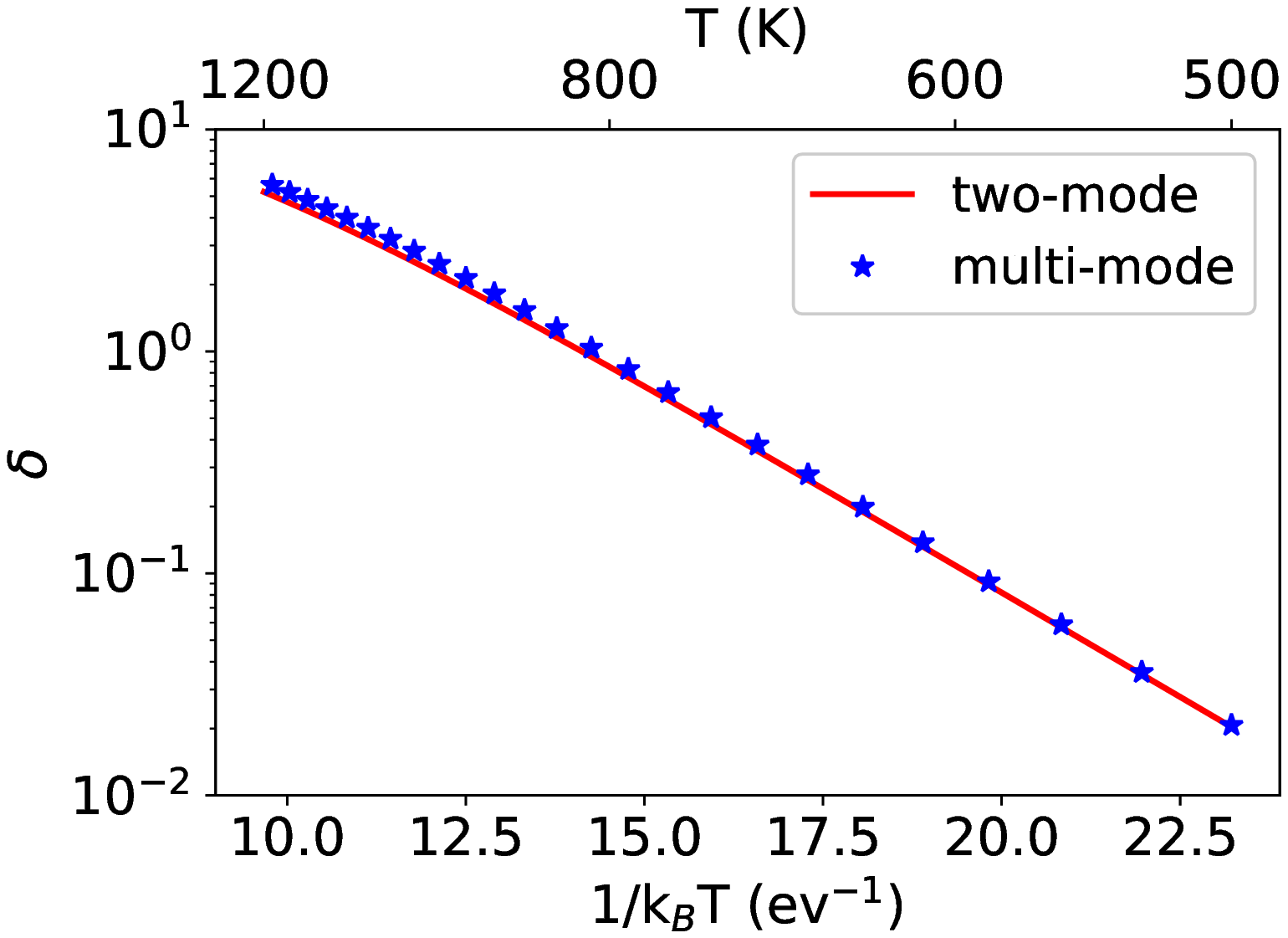}}
\caption{(a) Triple junction and GB (scaled) mobilities, $\tilde{M}_{TJ}$ and $\tilde{M}_{GB}$, versus temperature, comparing the two-mode and multi-mode predictions. (b) Arrhenius plots of the scaled TJ and GB multi-mode mobilities. This plot also shows the equilibrium disconnection densities for the primary and secondary disconnection modes (appropriately  shifted for comparison with the mobility data); these have formation energies of 0.20~eV and 0.64~eV, respectively. (c) The triple junction drag parameter $\delta$ versus temperature, comparing predictions based on two  and multiple modes.
}
\label{fig:Mobility}
\end{figure}

We compute the TJ mobility $M_{TJ}$, GB mobility $M_{GB}$ and TJ drag  parameter $\delta$ for the case of two disconnection modes (primary ($n=1, j=0$) and secondary  ($n=1, j=-1$) disconnections) and the case of multiple disconnection modes (all disconnection modes with $c_i/c_1 \geq 10^{-6}$) for this $\Sigma37$ GB.
The TJ mobility $M_{TJ}$ is obtained from the analytical expression Eq.~(\ref{eq:M_tj_two}) for two disconnection modes and by numerically solving the optimization problem for multiple disconnection modes.
The grain boundary mobility $M_{GB}$ and the TJ drag  parameter $\delta$ are then calculated based on Eqs.~(\ref{eq:M_gb}) and (\ref{eq:drag_delta}), respectively.
The simulation results are shown in Fig.~\ref{fig:Mobility} for the scaled, dimensionless quantities $\tilde{M}_{GB}={M}_{GB}/({M}_{d}c_1a_0L_0)$, $\tilde{M}_{TJ}=M_{TJ}/({M}_{d}c_1a_0)$ and $\delta=\tilde{M}_{TJ}/\tilde{M}_{GB}$.

Examination of Fig.~\ref{fig:Mobility} shows that the triple junction and GB mobilities increase with increasing  temperature.
The prediction of the mobilities based on the analytical expression for the two disconnection modes are almost identical to those  obtained by solving for multiple disconnection modes (using the numerical optimization approach).
This implies that accounting for the primary and secondary disconnections is sufficient for predicting the temperature variation of the GB and TJ mobilities (higher order modes have little effect), as discussed above.

The Arrhenius plots of $\tilde{M}_{GB}$ and $\tilde{M}_{TJ}$ (based on multiple disconnection modes)  and the equilibrium concentrations of primary and secondary disconnection modes (Fig.~\ref{fig:Mobility}b) show several interesting features.
First, the temperature dependence of the GB mobility is nearly identical with that of the equilibrium concentration of primary disconnection mode (except at high temperature).
This demonstrates that for this GB, GB migration is mainly controlled by a single disconnection mode.
In this case the formation energy of the secondary disconnection mode is $\sim3$ times that of the primary mode; the secondary mode may be important for GB in which the primary and secondary mode formation energies are more similar.
Second, the temperature dependence of the TJ mobility is nearly identical with that of the equilibrium concentration of \emph{secondary} disconnection mode.
This demonstrates that satisfying the zero TJ Burgers vector condition by activating two/multiple disconnection modes is necessary (and limiting) for TJ migration.
Since the GB and TJ mobilities both scale in an approximately Arrhenius manner with respect to temperature, the TJ drag parameter $\delta$ scales in an Arrhenius manner  with an activation energy equal to the difference between that of the TJ  and GB mobilities (which are nearly identical to the formation energies of the secondary and primary disconnection modes).
This is indeed seen in Fig.~\ref{fig:Mobility}(c).
The increase of the TJ drag parameter with increasing temperature indicates the transition from  TJ-controlled  to  GB-controlled grain growth with increasingly temperature ($\delta\ll1$ implies TJ drag-controlled grain growth;  $\delta\gg1$ implies GB migration-controlled grain growth), as has been widely observed in the experiments \cite{CZUBAYKO1998influence,Mattissen2005TJdrag}.
If this GB in Cu is viewed as ``typical'', this change in grain growth behavior should occur at $\sim800$ K (where $\delta \sim 1$).

\begin{figure}[!ht]
\centering
\subfloat[]{\includegraphics[width=0.33\textwidth]{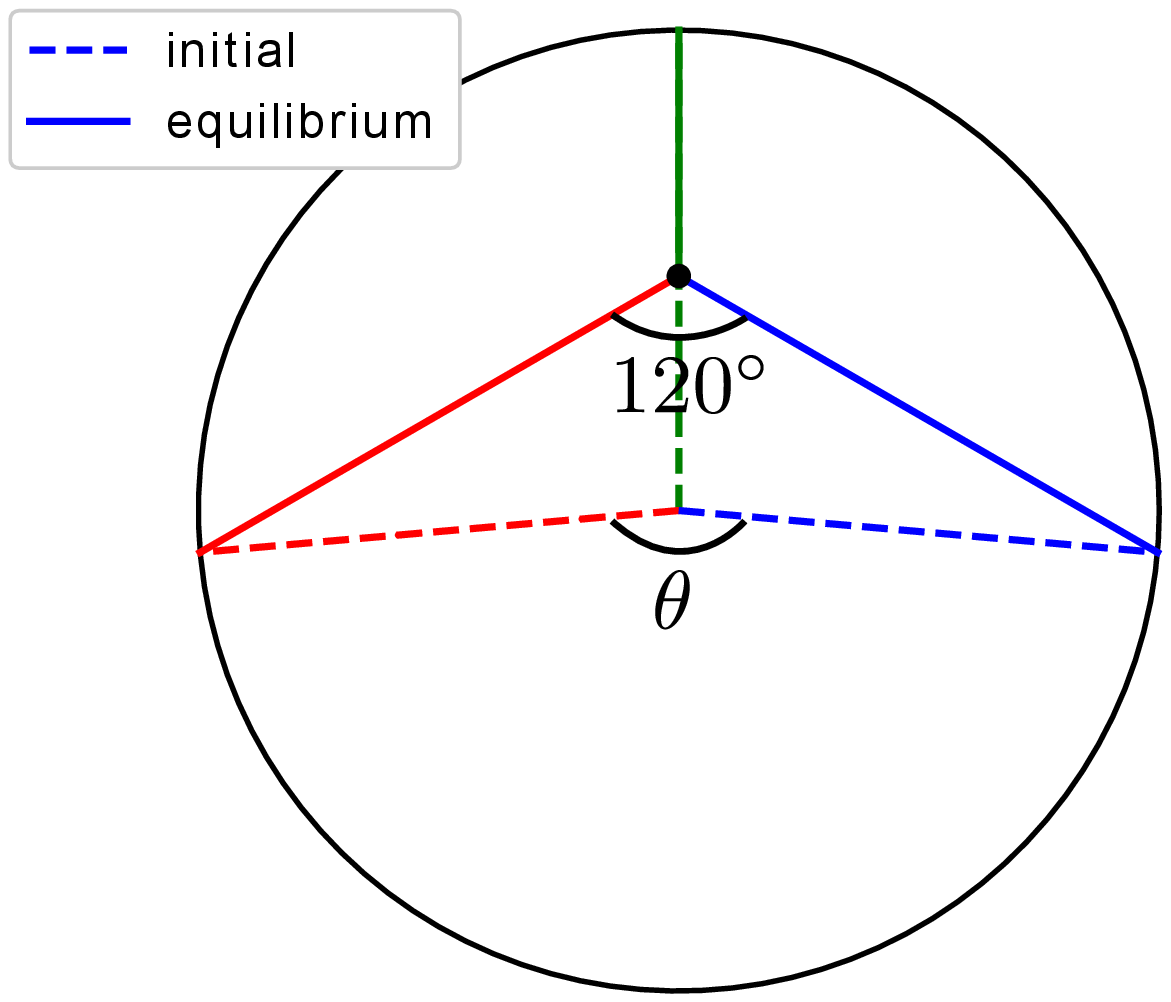}}
\subfloat[]{\includegraphics[width=0.33\textwidth]{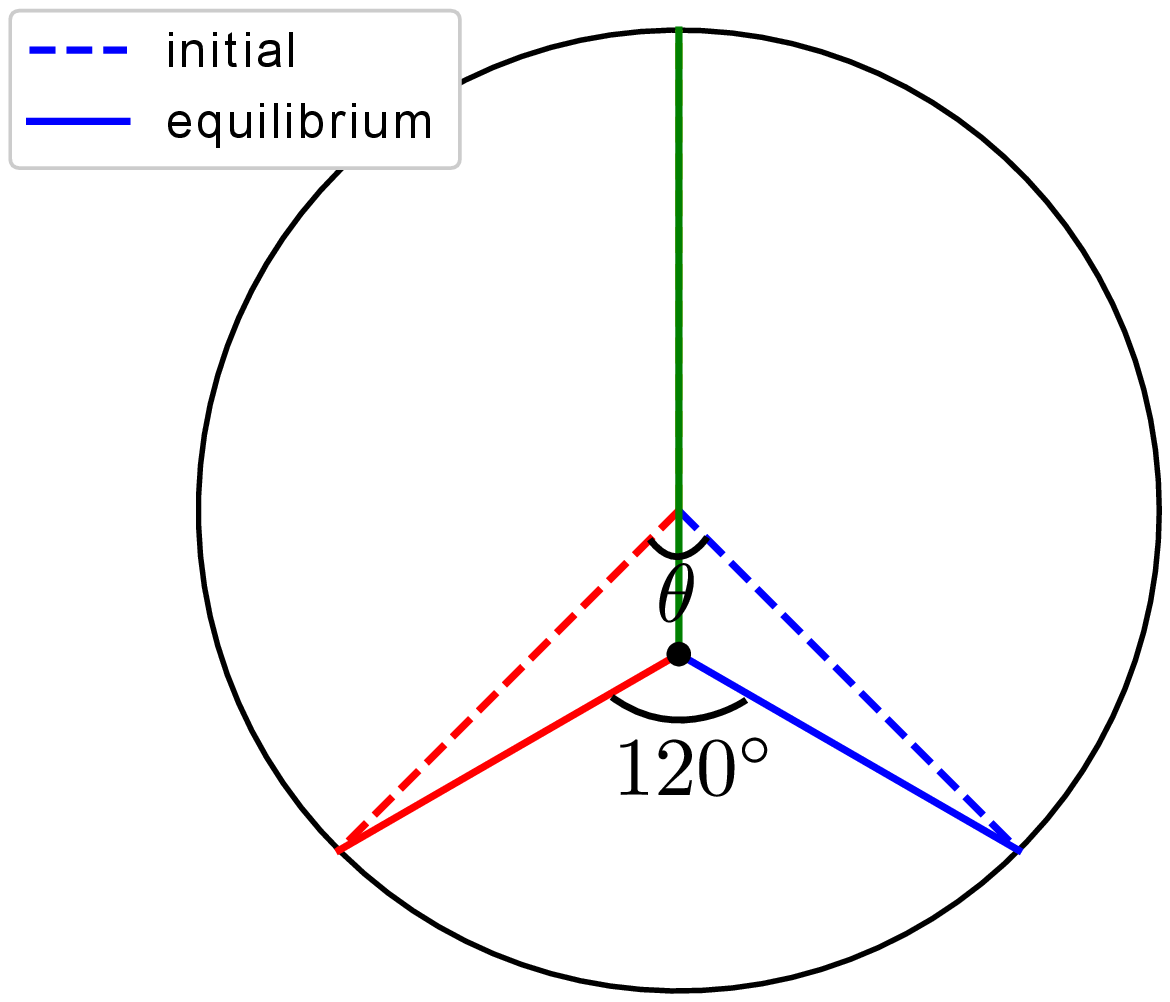}}
\subfloat[]{\includegraphics[width=0.33\textwidth]{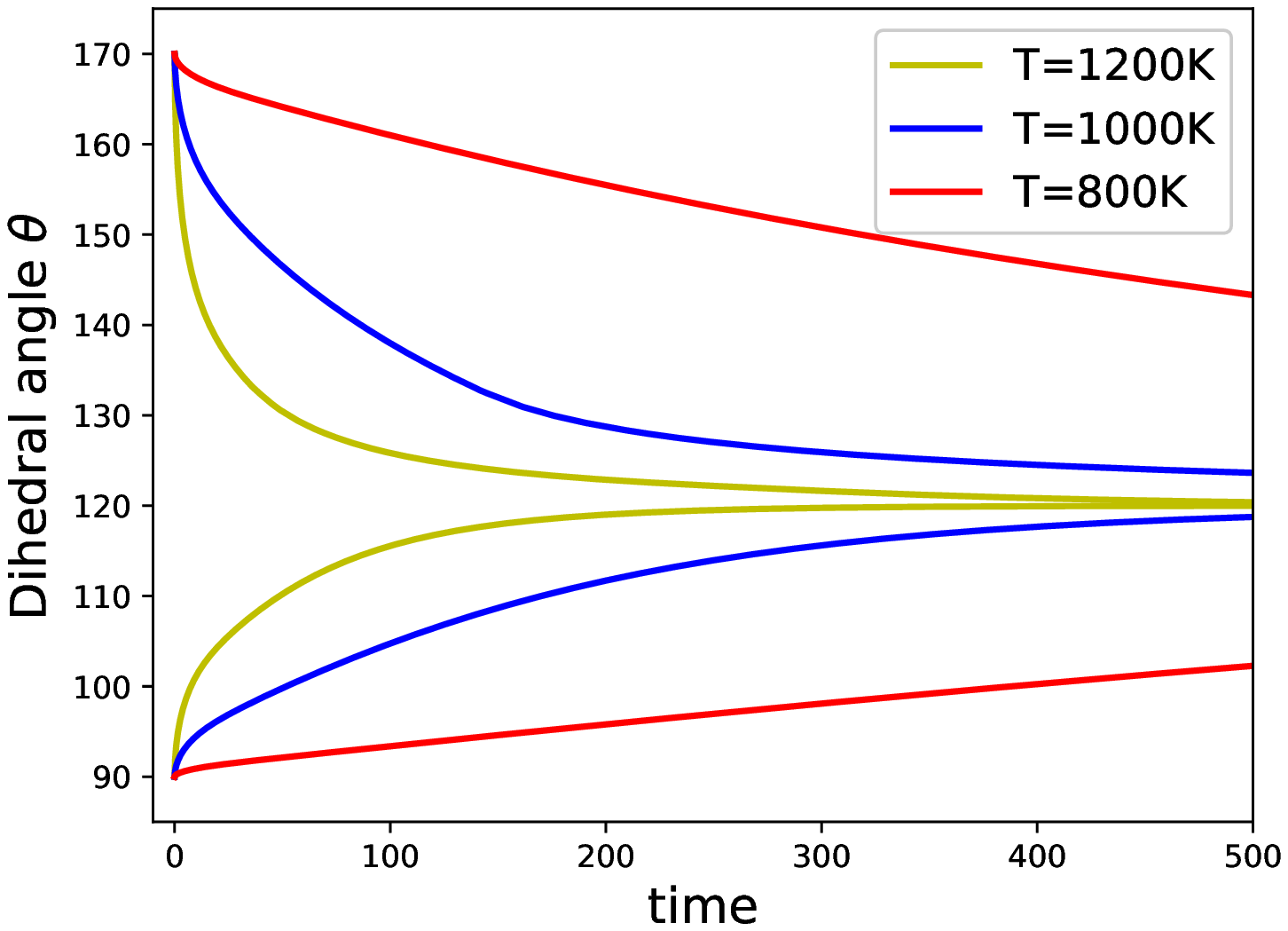}}
\caption{Triple junction and grain boundary migration in a tri-grain system. In (a) and (b), the initial configurations are shown with dashed lines and the equilibrium (final) configurations with solid lines for GBs pinned on the circle perimeters, as shown. In (a) the TJ moves up, while in (b) it moves down to establish the equilibrium TJ dihedral angles of 120$^\circ$.  (c) The time dependence of the dihedral angles for cases (a) (dihedral angle $\geq120^\circ$) and (b) (dihedral angle $\leq120^\circ$) for several temperatures. }
\label{fig:tj_motion}
\end{figure}

We also implemented our disconnection-based GT/TJ network model in continuum simulations for the dynamics of a tri-grain system for the symmetric profile case discussed in Sec.~\ref{subsec:M_tj} and shown in Figs.~\ref{fig:tj_motion} (a)/(b).
While the TJ is free to migrate in these simulations, the opposite ends of the  three GBs are pinned to fixed points on a circle of radius 100 \AA.
Molecular dynamics simulations were previously reported for the same geometry \cite{thomas2019TJ}.
In Figs.~\ref{fig:tj_motion} (a)/(b), the TJ moves up/down and the effective dihedral angle between the red and blue GBs approach the equilibrium angle dihedral angle of $120^\circ$ from above/below.
The effective dihedral angle is defined here as the angle at the apex of the isosceles triangle drawn  between the fixed ends of the red and blue GBs and the TJ.
In both cases (a) and (b), the effective dihedral angle relaxes to the equilibrium dihedral angle faster with increasing temperature.
These results show a very good qualitative agreement with the MD simulation results in \cite{thomas2019TJ}.

\section{Summary and discussion}\label{sec:summary}
We provide a unified continuum model for coupled  grain boundary and triple junction migration based on the microscopic mechanism of GB migration.
The microscopic mechanism involves the migration of line defects (disconnection) along the GB that have both step and dislocation character.
The grain boundary and triple junction migration is described in terms of the thermally-activated nucleation and kinetically-limited motion of disconnections of multiple types/modes (the possible disconnection modes are set by the crystallography).
The governing equations of the evolution of the GB/TJ system are derived within a variational framework based on the principle of maximum dissipation of energy (i.e., an Onsager variational approach), that includes local GB energy, long-range elastic interactions between disconnections, applied stresses, and jumps in chemical potential across GBs. The GB motion and the ``apparent" GB mobility are determined by the formation of and competition between  all disconnection modes in response to all of these driving forces.
The TJ motion is driven by the capillary and elastic forces (but not chemical potential jumps) subject to  constraints arising from the underlying disconnection mechanism.
The resultant TJ dynamics is modeled as an optimization problem with respect to the disconnection flux, subject to geometric constraints, conservation of Burgers vector, and thermal-kinetic limitations on the disconnection flux.
Both GB and TJ mobilities are determined by the more fundamental disconnection  properties, as well as the GB/TJ network geometry and the source of the driving forces, rather than being intrinsic properties.
We perform both analysis and numerical computations to determine GB/TJ mobilities and the TJ drag effect.
The results demonstrate that while the apparent GB mobility is most commonly controlled by the formation of the primary (lowest formation energy) disconnection mode, it is the secondary  (second lowest formation energy) disconnection mode that controls triple junction mobility.
This difference between which disconnection mode dominates the two types of mobilities naturally gives rise to TJ drag such that at low temperature TJ drag controls grain growth while at high temperature grain growth is controlled by the migration of the GBs.

In  conventional continuum models for the evolution of polycrystalline microstructures, the motion of GBs is modeled as the evolution of continuum surfaces.
Most commonly, such surfaces are assumed to evolve via GB mean curvature flow.
Such models do not include such important effects as the role of crystallography, stress or temperature other than through empirical parameters.
While atomic-scale  models provide complete information on GB crystallography, stress, and temperature through the evolution of all of the atoms in the microstructure, such models are severely limited by  the unaffordable computational cost associated with the characteristic  length and time scales of grain growth.
The present model for  GB/TJ dynamics is in fact a mesoscale model that resides between an atomic-scale and a continuum model, that accounts for crystallography, stress and temperature effects absent in the continuum approach but at a computational cost much more reasonable than the atomistic approach.
In our model, GB and TJ mobilities are not external parameters, but rather the result of the fundamental disconnection properties and the geometry of the microstructure.

The proposed continuum framework can be applied for describing the kinetics of polycrystalline microstructure in a manner that is consistent with the underlying disconnection dynamics.
The disconnection model provides a unified mechanism for GB and TJ migration and can simultaneously account for such phenomena as grain boundary sliding and grain rotation \cite{han2018}.
Since the disconnection model is consistent with dislocation models for plasticity within grains, this approach provides the opportunity for linking continuum descriptions of microstructure evolution with continuum descriptions of plastic deformation. Further developments are required to implement the present disconnection-based continuum model into large scale simulations of microstructure evolution; e.g., through integration of this model with front-tracking, implicit sharp-interface or diffuse interface methods.
Such models would require the inclusion of internal state variables (disconnection densities) along the interfaces as well as the evolution of these local state variables.

\section*{Acknowledgments}
C.W. and Y.X. acknowledge support from the Hong Kong Research Grants Council General Research Fund 16302818.
The research contributions of D.J.S. and J.H. were sponsored by the Army Research Office and were accomplished under Grant Number W911NF-19-1-0263. The views and conclusions contained in this document are those of the authors and should not be interpreted as representing the official policies, either expressed or implied, of the Army Research Office or the U.S. Government. The U.S. Government is authorized to reproduce and distribute reprints for Government purposes notwithstanding any copyright notation herein.

\bibliographystyle{siamplain}
\bibliography{myreference_abbr}
\end{document}